\documentclass[12pt]{iopart}

\expandafter\let\csname equation*\endcsname\relax
\expandafter\let\csname endequation*\endcsname\relax

\usepackage{amsmath,amsfonts,amssymb}
\usepackage{caption}
\usepackage{subcaption}

\usepackage{graphicx,psfrag}

\renewcommand{\rmd}{\mathrm{d}}

\newcommand{\laplace}{\triangle}

\newcommand{\energy}{\mathcal{E}}

\begin{document}

\title{Energy Spreading in Strongly Nonlinear Disordered Lattices}

\author{M.~Mulansky and A.~Pikovsky}
\address{Department of Physics and Astronomy, Potsdam University,
  Karl-Liebknecht-Str 24, D-14476, Potsdam-Golm, Germany
}
\eads{\mailto{mulansky@uni-potsdam.de},\mailto{pikovsky@uni-potsdam.de}}
\date{\today}

\begin{abstract}
We study scaling properties of energy
spreading in disordered strongly nonlinear Hamiltonian lattices. Such lattices consist of nonlinearly coupled local linear or nonlinear oscillators, and demonstrate a rather slow, subdiffusive spreading of initially localized wave packets.
We use a fractional nonlinear diffusion equation as a heuristic model of this process, and confirm that the scaling predictions resulting from a self-similar solution of this equation are indeed applicable to all studied cases. We show that the spreading in nonlinearly coupled linear oscillators slows down compared to a pure power law, while for nonlinear local oscillators a power law is valid in the whole studied range of parameters.

%\keywords{Weak chaos \and localization \and scaling}
\end{abstract}
\submitto{NJP}
%\tableofcontents
\maketitle

\section{Introduction}

The general understanding of the relation between chaos in classical systems, and ergodicity and thermalization is still far from complete nowadays.
Intuitively, one expects from high-dimensional, non-integrable complex systems to demonstrate strong chaos and thus it seems reasonable to expect thermalization.
This is essentially the fundamental assumption of classical thermodynamics~\cite{Landau-lifshitz-80}.
The conditions under which this assumption can be safely made, however, is still an open question.
It is not known what level of ``chaoticity'' or ``complexity'' is required to ensure thermalizing behavior.
Chaos can be often seen as a consequence of nonlinear perturbations of an integrable system.
The solutions of the unperturbed, integrable part of such a system are called modes.
Starting with only a few initially excited modes, one can view thermalization
as spreading in the mode-space, i.e. the excitation of new modes, due to the nonlinear chaotic interactions.

However, already the first attempts to follow such a thermalization of modes initiated by Fermi, Pasta, and Ulam revealed many extremely nontrivial effects, still not completely understood (see  Refs.~\cite{Gallavotti-08,Chaos-fpu-05} for recent progress of the FPU problem). 
A very important case is when the integrable modes are spatially localized.
Then the thermalization process is a spatial diffusion where more and more modes get excited.
This allows to connect the rather abstract concept of thermalization in the mode-space with the very intuitive phenomenon of spatial diffusion.
A prominent example where this has been studied very extensively in the past, is the interplay between nonlinearity and disorder.
In this case, due to Anderson localization, linear eigenmodes are exponentially localized and the spectrum is purely discrete~\cite{Abrahams-10}. 
Recent numerical experiments with nonlinear disordered lattices have demonstrated that the initially localized wave packets spread in a very weak, subdiffusive manner~\cite{Shepelyansky-93,%
Molina-98,%
Pikovsky-Shepelyansky-08,Fishman-Krivolapov-Soffer-08,%
Garcia-Mata-Shepelyansky-09,%
Flach-Krimer-Skokos-09,Skokos_etal-09,%
Mulansky-Ahnert-Pikovsky-Shepelyansky-09,%
Skokos-Flach-10,Flach-10,Laptyeva-etal-10,%
Mulansky-Pikovsky-10,Johansson-Kopidakis-Aubry-10,Mulansky-Ahnert-Pikovsky-11,%
Ivanchenko-Laaptyeva-Flach-11,Fishman-Krivolapov-Soffer-12}.
A complete theoretical understanding of the subdiffusive behavior has not been presented, but it is mostly agreed on that the spreading in these models is induced by weak chaos.
However, the true asymptotic behavior is still discussed in some of these models with recent claims that spreading might stop due to an extinction of chaos~\cite{Pikovsky-Fishman-11,Johansson-Kopidakis-Aubry-10,Roy-Pikovsky-12}.

In this paper we follow the scaling approach to this problem, first formulated in \cite{Mulansky-Ahnert-Pikovsky-11} and recently extended to two-dimensional systems~\cite{Mulansky-Pikovsky-12}. 
We will try to establish and to check numerically the scaling relations for the properties of spreading, in dependence on the total energy of the initial wave packet. 
In these works, the nonlinear diffusion equation (NDE) was proposed to describe the spreading process and this assumption was verified by several numerical simulations.
Here, we will generalize this model by introducing the fractional nonlinear diffusion equation (FNDE), and we will present new numerical results that will show that in some cases indeed only the FNDE gives a correct scaling
description of the spreading process.
We formulate the scaling relations based on this equation, and check their validity for nonlinear lattices.

We will start with formulating the object of our study, strongly nonlinear Hamiltonian lattices. 
We present a phenomenology of energy spreading and define the statistical quantities characterizing it in section \ref{sec:snl}. 
Next, we introduce a phenomenological model that we use to describe the properties of the spreading process, namely the fractional nonlinear diffusion equation (section \ref{sec:nde}). 
From its scaling properties, we derive spreading predictions for the strongly nonlinear Hamiltonian lattices.
In section~\ref{sec:nres} we  present extensive numerical calculations for different classes of the nonlinearity.
These results are compared with the predictions from the FNDE and we identify different degrees of confirmation for different nonlinear classes.
We end with concluding remarks where the found ``universality classes'' are summarized.

\section{Strongly nonlinear lattices}
\label{sec:snl}

The main goal of this paper is to study properties of energy spreading in strongly nonlinear lattices. 
By strongly nonlinear we understand lattices where the coupling is described by nonlinear functions that disappear in the linear limit. 
So there are no linear waves (phonons) in such lattices and energy transport can solely be induced by the nonlinear coupling. 
Such lattices can be introduced in the framework of equations for the complex amplitudes (and then one obtains a strongly nonlinear generalization of the nonlinear Schr\"odinger equation) or as a generalization of a Hamiltonian Klein-Gordon lattice. 
In this work we follow the latter way. 
Moreover, we restrict ourselves to pure power-law nonlinearities.

\subsection{Hamiltonian} \label{sec:1d_model}
In one dimension we formulate a strongly nonlinear lattice 
in terms of a Hamilton function for positions~$q_k$ and momenta~$p_k$ of oscillators labelled by site index $k$:
\begin{equation} 
 H = \sum_k \frac{p_k^2}{2} + W\frac{\omega_k^2}{\kappa} q_k^\kappa + \frac\alpha\lambda (q_{k+1} - q_k)^\lambda\;.
\label{eqn:general_hamiltonian}\end{equation}
Here $\kappa \geq 2$ and $\lambda > 2$ denote the powers of the on-site potential and the coupling term, respectively. For $\kappa=2$ we have a chain of nonlinearly coupled linear on-site oscillators; for $\kappa>2$ the on-site oscillators are nonlinear as well. Below we study situations with and without disorder, the latter is introduced via the variations of the parameters of the local potential
$\omega_k$ (these are linear frequences of the oscillators if $\kappa=2$ and parameters of the nonlinear on-site potential if $\kappa>2$).
%We will consider three cases:
%(i) no disorder, $\omega_k=1$; (ii) ``soft'' disorder, in this case 
%the ``frequencies'' $\omega_k$ are chosen iid. from $[0,1]$; 
%(iii) ``hard'' disorder, here  the ``frequencies'' $\omega_k$ 
%are chosen iid. from $[0.5,1.5]$.
Note, that the integrable part of this system are uncoupled oscillators ($\alpha\rightarrow0$), which means that the modes of the integrable system are extremely localized on one site.

\subsection{Rescaling}
\label{sec:resc}
Hamiltonian (\ref{eqn:general_hamiltonian})  contains two parameters $W$ and $\alpha$ that determine the time scale and the ratio of local to coupling potentials. For different local and coupling nonlinearities, i.e. for $\kappa\neq \lambda$, we can get rid of these two parameters by rescaling the canonical variables and time as follows:
\begin{equation} \label{eqn:asymmetric_scaling}
q_k  \to W^b \alpha^{-b} q_k\;, \quad
p_k \to W^{\lambda b/2}\alpha^{-\kappa b/2} p_k\;, \quad
 t   \to W^{(2-\lambda)b/2}\alpha^{(\kappa-2)b/2} t\;,
\end{equation}
with $b = 1/(\lambda-\kappa)$. $W$ and $\alpha$ disappear from the equations and we are left with the total energy $\energy$ as the only relevant parameter depending on the initial state. Additionally, the distribution of local ``frequencies'' $\omega_k$ is relevant, while the width of this distribution is rescaled together with $W$. We will consider three cases:
(i) no disorder, $\omega_k=1$; (ii) ``soft'' local oscillators, in this case the ``frequencies'' $\omega_k$ are chosen iid. from $[0,1]$; (iii) ``hard'' local disorder, here  the ``frequencies'' $\omega_k$ are chosen iid. from $[0.5,1.5]$.
In the rescaled coordinates the Hamiltonian reads
\begin{equation} \label{eqn:ghr}
 H = \sum_k \frac{p_k^2}{2} + \frac{\omega_k^2 q_k^\kappa}{\kappa}  + \frac {(q_{k+1} - q_k)^\lambda}{\lambda}\;.
\end{equation}

One special and highly interesting case occurs when  
 the on-site and coupling terms have the same nonlinearity $\kappa = \lambda$.
As now all terms in  $q$ have the same power, one cannot set both parameters $W$ and $\alpha$ to one by rescaling as before. Instead, one can use the remaining freedom to set the total energy $\energy$ to unity: 
\begin{equation} \label{eqn:symmetric_scaling}
 q \to \energy^{1/\kappa}W^{-1/\kappa} q,\quad
 p  \to\energy^{1/2} p,\quad
 t \to  W^{-1/\kappa}\energy^{1/\kappa-1/2} \;.
\end{equation}
Particularly, this means that the energy is not a free parameter of the system but can rather be scaled to, say, $\energy = 1$, what also involves an
appropriate change of the time scale.
We note that the only remaining parameter is the ratio of strengths of on-site and coupling terms $\beta=\alpha/W$. The rescaled Hamiltonian now reads
\begin{equation} \label{eqn:ghrk}
 H = \sum_k \frac{p_k^2}{2} + \frac{\omega_k^2 q_k^\kappa}{\kappa}  + \beta\frac {(q_{k+1} - q_k)^\kappa}{\kappa}\;.
\end{equation}

\subsection{Phenomenology of energy spreading}

For the Hamiltonian systems (\ref{eqn:general_hamiltonian})  we state the following question: How does an initially localized field spread over the lattice? 
We focus on very large systems, where boundary effects are not so important (we will discuss their relevance in some cases below). The distribution of energy is characterized with its density
\begin{equation}
w_k= \frac{E_k}{\energy}=\energy^{-1}\left( \frac{p_k^2}{2} + W\frac{\omega_k^2}{\kappa} q_k^\kappa + \frac\alpha{2\lambda} [(q_{k+1} - q_k)^\lambda + (q_{k} - q_{k-1})^\lambda]\right)\;.
\end{equation}
We start typically with non-zero values of $w_k$ in a small interval (in most runs 10 sites), by chosing initial momenta from a Gaussian distribution, and follow the distribution $w_k(t)$ in time. 

\begin{figure}[t]
 %\psfrag{xlabel1}[cc][cc]{\footnotesize $\log_{10}L$}
 %\psfrag{ylabel1}[cc][cc]{\footnotesize $\log_{10}\Delta T$}
 \psfrag{xlabel2}[cc][cc]{\footnotesize \footnotesize lattice site $k$}
 \psfrag{ylabel2}[cc][cc]{\footnotesize time $t$}
 \psfrag{xlabel1}[cc][cc]{\footnotesize lattice site $k$}
 \psfrag{ylabel1}[cc][cc]{\footnotesize time $t$}
 \psfrag{sp-ti_46}[cc][cc]{\footnotesize }
 \psfrag{sp-ti_46_dis}[cc][cc]{\footnotesize }
% \begin{tabular}{cc}
\begin{subfigure}[b]{0.48\textwidth}
 \centering
 \includegraphics[width=\textwidth]{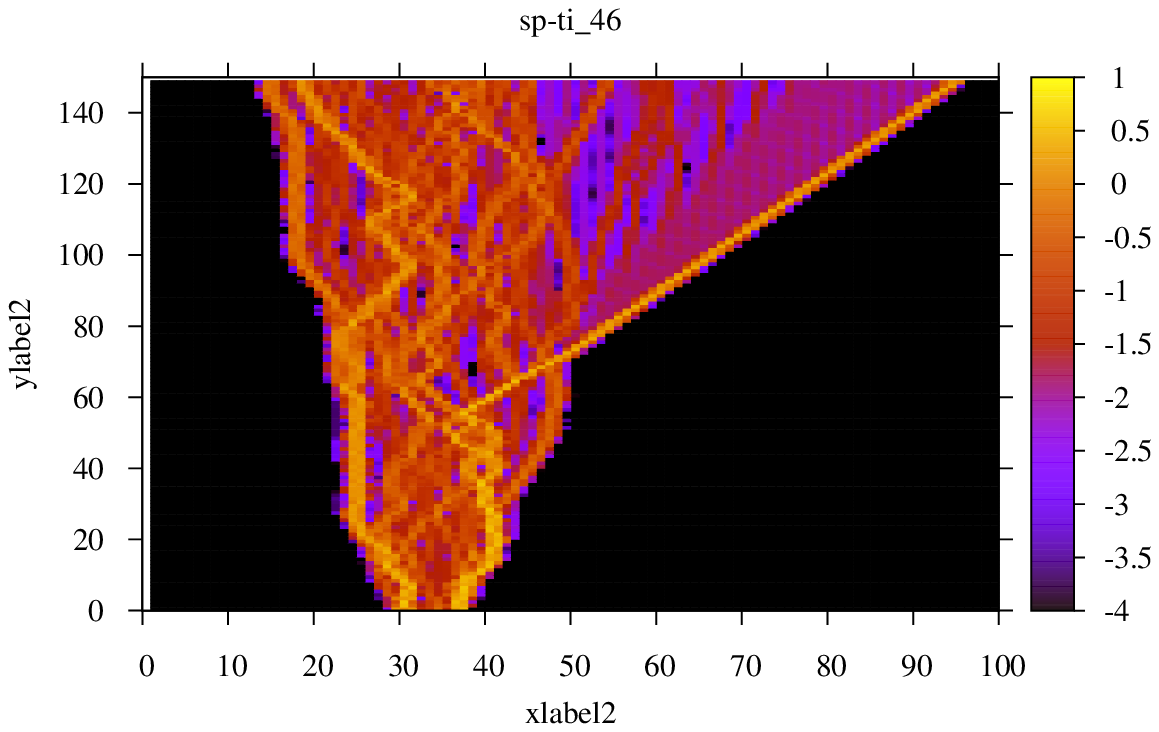}
 \caption{(a) regular lattice $\omega_k=1$.}
 \label{fig:4_6_re_wf}
\end{subfigure}
\begin{subfigure}[b]{0.48\textwidth}
 \centering
 \includegraphics[width=\textwidth]{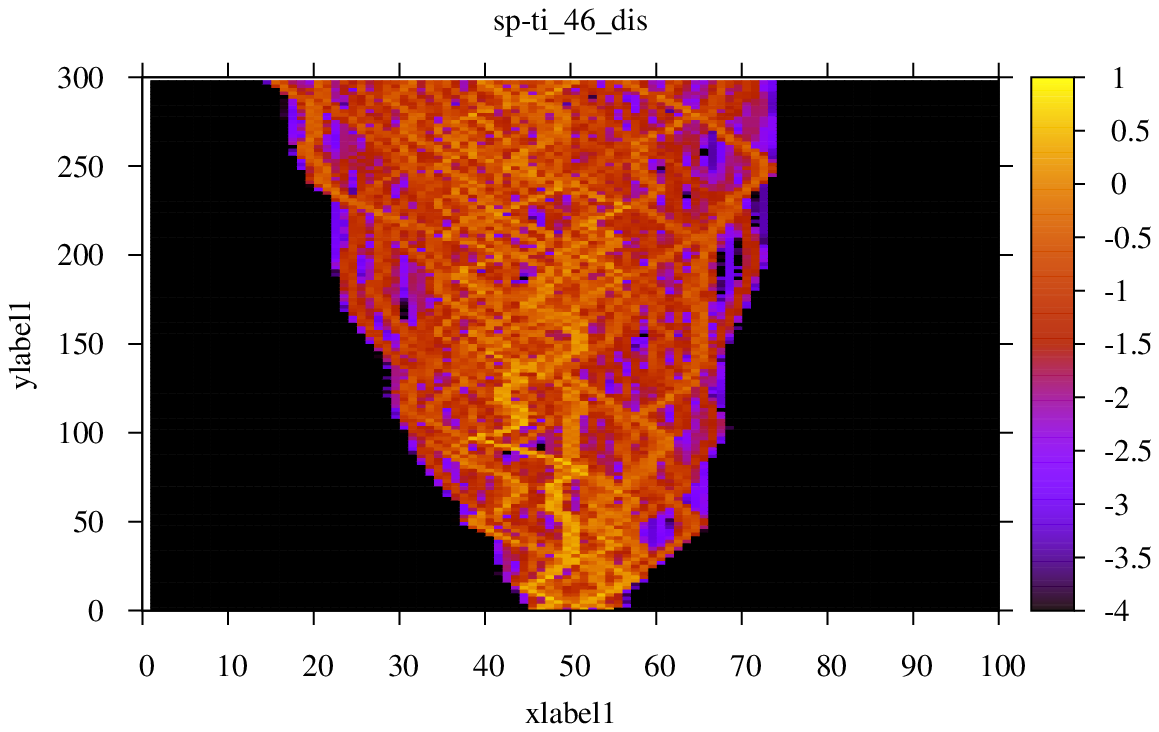}
 \caption{(b) hard disorder $\omega_k\in[0.5,1.5]$.}
 \label{fig:4_6_hd_wf}
\end{subfigure}

% \end{tabular}
 % self_sim.eps: 0x0 pixel, -2147483648dpi, 0.00x0.00 cm, bb=
 \caption{Time evolution of an initially localized state for $\kappa=4$, $\lambda=6$, $W=\beta=1$ and energy $\energy=10$. (a): regular lattice, (b): lattice with ``hard'' disorder. The color coding corresponds to the logarithm of the local energy excitation $\log_{10} w_k$. The initial excitation was uniform on 10 sites.
 }
 \label{fig:4_6_wf}
\end{figure}

In the case of a lattice without disorder ($\omega_k=1$), regular waves can propagate along the lattice. 
Such localized solitary waves -- compactons -- have been thoroughly studied in \cite{Ahnert-Pikovsky-09} for a lattice with $W=0$ (i.e. without local potential). 
An initially localized perturbation emits compactons that dominate the process of energy spreading. 
For $W\neq 0$ it is not known if exact compactons exist in such lattices. 
In numerics, we quite often observe ``quasi-compactons'' that propagate ballistically over large distances but lose energy and therefore eventually stop. 
We illustrate this in Fig.~\ref{fig:4_6_re_wf}. 
In panel~\ref{fig:4_6_hd_wf} we show the same initial conditions in a disordered lattice, here the propagation of ``quasi-compactons'' is blocked by disorder and one observes a slow spreading of the wavepacket, which will later be quantified as subdiffusive.

Additionally to the observation of ``quasi-compactons'', we see that in disordered strongly nonlinear lattices at any finite time the distribution of energies is strongly localized, and has sharp edges (we expect that generally the field at the edges decays superexponentially fast, as for breather solutions in such lattices~\cite{Rosenau-Schochet-07}). 
This sharpeness is illustrated in Fig.~\ref{fig:2_4_wf}. 
%For regular lattices, a large part of energy remains localized, while small ``tails'' due to propagating quasi-compactons are not excluded; the field typically does not have sharp edges. Moreover, we cannot exclude that ``quasi-compactons'' can reach the boundaries of the lattice in numerical simulations. (For disordered lattices, due to very sharp edges, we increase the domain in accordance with the propagation).  

\subsection{Measures of spreading}
\label{sec:ms}

In a statistical context, the spread of a distribution $w_k$ can be quantified via entropies, most suitable are the R{\'e}nyi entropies:
\[
I_q=\frac{1}{1-q}\ln\sum_k w_k^q,
\]
that allow one to characterize also the spikeness/flattness of the distribution (in the context of energy spreading in disordered lattices this approach was introduced in~\cite{Mulansky-Pikovsky-10}). 
We restrict here to the the entropies $I_1$ and $I_2$, which are nothing else than the usual Boltzmann entropy and the logarithm of the participation number $P$:
\begin{equation}\label{eq:entrdef}
I_1=-\sum w_k\ln w_k\qquad I_2=-\ln\sum_k w_k^2=\ln P\;.
\end{equation}
The participation number is a characteristic of the width of the wave packet rather popular in the context of Anderson localization studies~\cite{Mulansky-Pikovsky-10,Flach-Krimer-Skokos-09,%
Veksler-Krivolapov-Fishman-09}. Both entropies define the effective width of the wave packet as $\mathcal{L}_{1,2}=\exp(I_{1,2})$ (in particular, $\mathcal{L}_2=P$). As individual dependencies $\mathcal{L}(t)$ demonstrate enormous fluctuations, we perform an averaging of the entropies $I_{1,2}(t)$ over many realizations of disorder, thus obtaining smoothly growing widths $  \mathcal{L}(t)$. 

\begin{figure}[tb]
 \begin{subfigure}[b]{0.5\textwidth}
  \includegraphics[width=\textwidth]{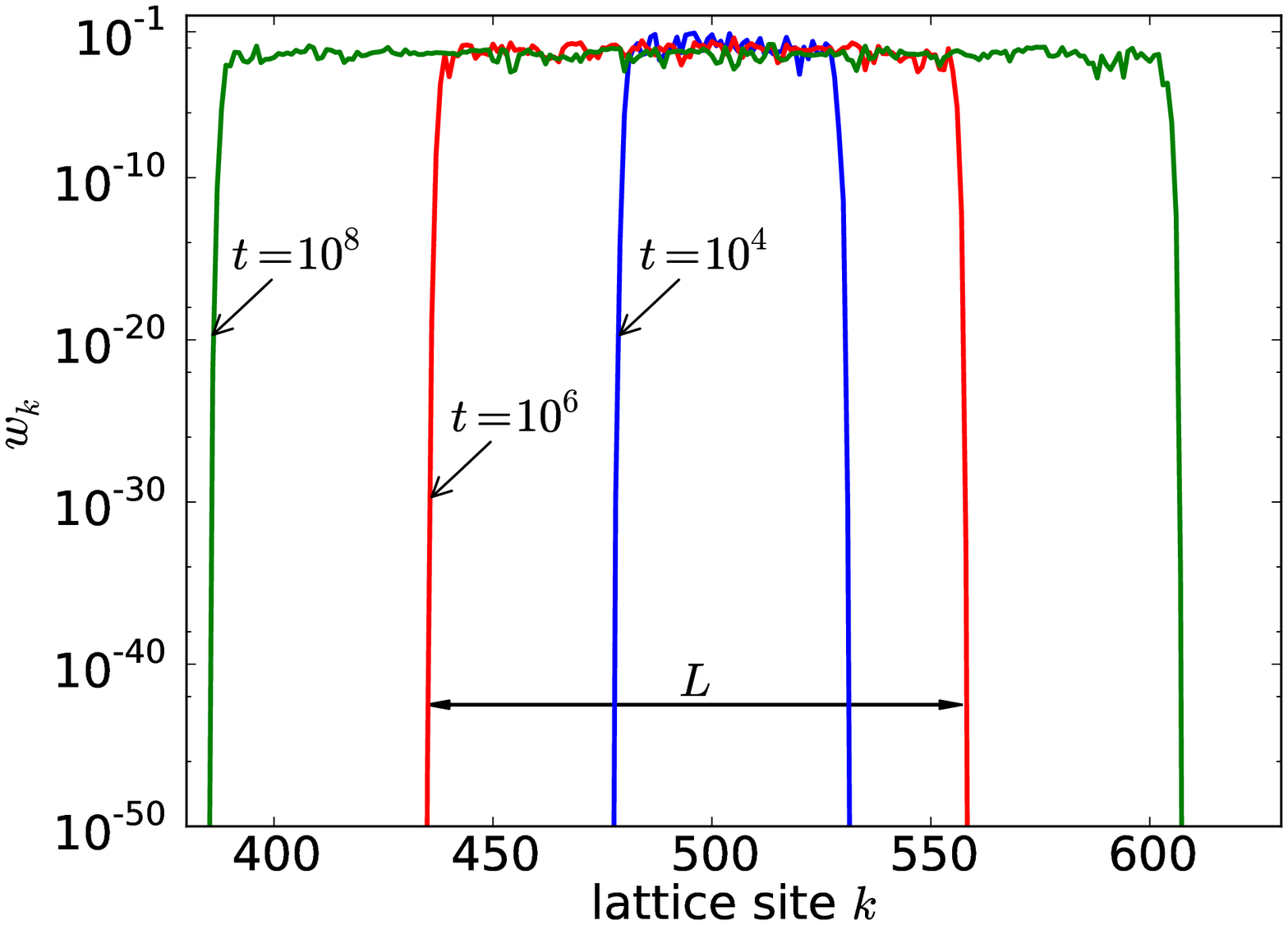}
  \caption{(a) Spreading of a single site excitation.}
  \label{fig:2_4_wf}
 \end{subfigure} \hfill
 \begin{subfigure}[b]{0.45\textwidth}
  \includegraphics[width=0.95\textwidth]{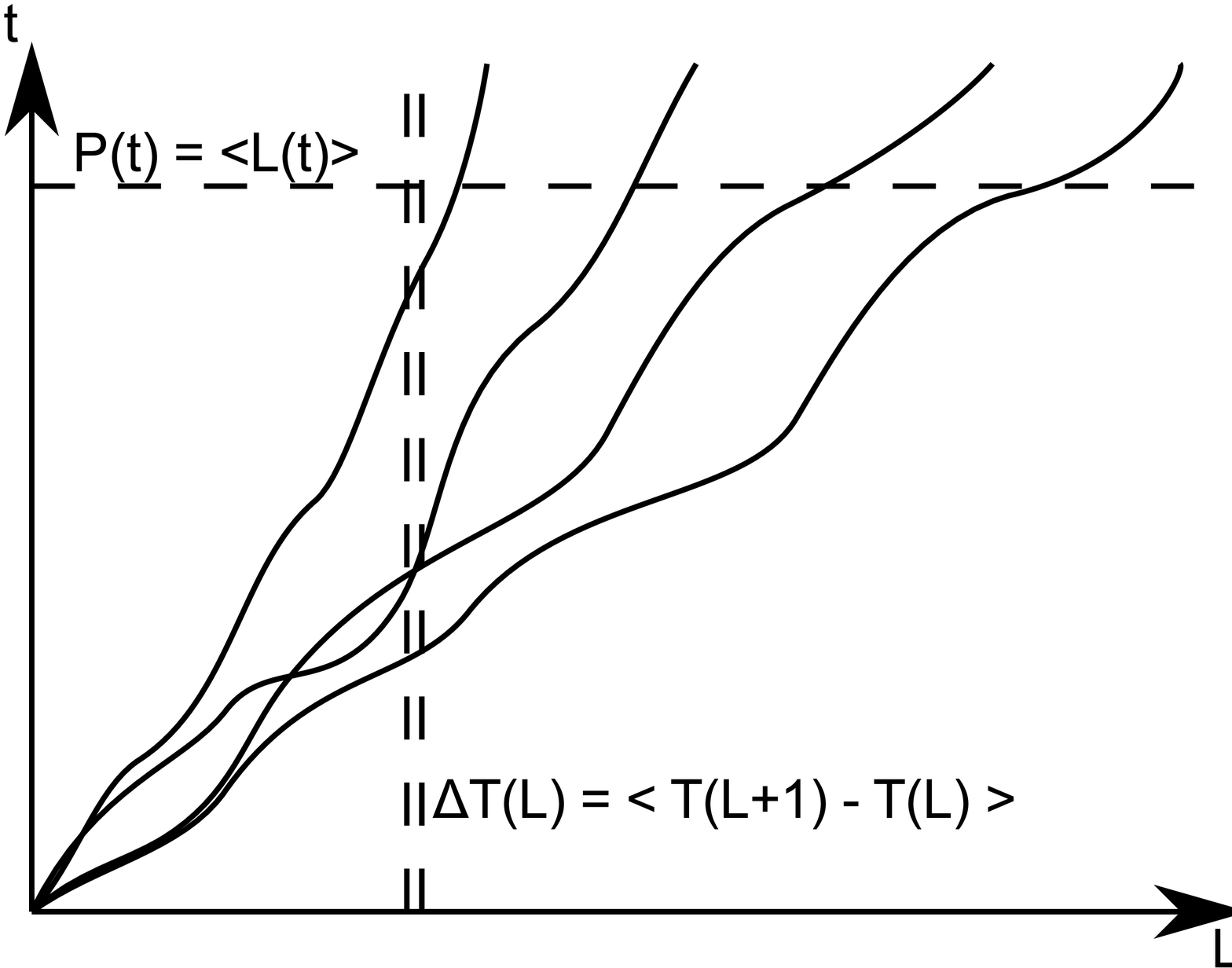}
  \caption{(b) Different ways to measure spreading.}
  \label{fig:exc_time_spreading}
 \end{subfigure}
 \caption{Panel (a) shows the spreading of an initial single site excitation for $\kappa=4$, $\lambda=6$ and energy $\energy=1.0$. The plot shows local energy $w_k$ vs.\ lattice site $k$ for increasing times $10^4$, $10^6$, $10^8$ (inner to outer curves). Note the logarithmic scaling of $w_k$ and the exponential drops in this scale. In panel (b) we schematically plot the two ways to measure spreading. Averaging at fixed $L$ means averaging at fixed energy density, contrary to averaging at fixed time.
 }
\end{figure}

For the strongly nonlinear lattices another approach~\cite{Mulansky-Ahnert-Pikovsky-11} is even superior to the calculation of the entropies. Here we determine the width $L$ of the wave packet as the distance between its sharp 
edges as seen in Fig.~\ref{fig:2_4_wf} (independently on the distribution of the energy between these edges). After determining the spatial extend $L$, we measure the time $\delta T$ required to excite one new lattice site.
So suppose we have $L$ lattice sites being excited, then $\delta T(L)$ is the time required to pass from $L$ to $L+1$ lattice sites (so this quantity is in fact a first passage time).
We define a lattice site as excited when its local energy exceeds some border $E_B = 10^{-50}$.
The actual value of $E_B$ was chosen arbitrarily, but any other value, e.g.~$E_B=10^{-100}$ would produce similar results. The quantity $\delta T$ can be interpreted as a propagation time for $L\to L+1$, it can be determined for each particular realization of disorder and initial condition. After having the ensemble of these propagation times at given $L$, we calculate the average propagation time $\Delta T$ as the geometric average of $\delta T$ (equivalently, we average the logarithms $\Delta T=\exp[\langle\log\delta T\rangle]$). 
%$\Delta T$ can be understood as the time to advance one lattice site $\Delta X=1$, hence $\Delta T$ is an approximation for the inverse spreading velocity $\rmd t / \rmd X$.

This second approach is superior to the measurement of the effective spatial extent $\mathcal{L}$ for two main reasons:\\
(i) First, $\Delta T$ has no explicit time dependence, hence any prehistory 
does not appear in later measurements of $\Delta T$.
It is therefore easier to compare different realizations and simulations for different parameter values and different initial conditions using $\Delta T$.\\
(ii) The second advantage  relates to the procedure of
averaging over many realizations of the spreading trajectories.
By averaging $\Delta T(L)$ for fixed $L$, we average over situations with the same energy density $w = \energy/L$.
If, in contrast, different realizations of $\mathcal{L}(t)$ are averaged for a fixed time $t$, situations with different energy densities $w$ are averaged together, which is not reasonable if the density $w$ is the crucial parameter on which the properties of the propagation should depend.
This is schematically sketched in Fig.~\ref{fig:exc_time_spreading}.

\section{Fractional nonlinear diffusion equation}
\label{sec:nde}
We study spreading which is induced by nonlinear chaotic interactions between oscillators, it disappears in the integrable (linear) limit. 
It is known for Hamiltonian systems that chaos might lead to diffusive behavior which can be understood as the result of ``intrinsic stochasticity'' induced by the chaotic motion~\cite{Zaslavsky72,Lichtenberg-Lieberman-92}.
In former works, the nonlinear diffusion equation (NDE) was introduced and remarkable similarities between its scaling properties and the spreading behavior and numerical results for strongly nonlinear lattices were found~\cite{Mulansky-Ahnert-Pikovsky-11,Mulansky-Pikovsky-12,Mulansky_phd}.
The nonlinear diffusion equation describes the spatio-temporal evolution of a density $\rho(x,t)$ with a density dependent diffusion ``constant'' $D(\rho)\sim\rho^a$:
\begin{equation}  \label{eqn:NDE}
 \frac{\partial \rho}{\partial t} = D_0 \frac\partial{\partial x} \left(\rho^a \frac{\partial \rho}{\partial x}\right) = \frac{D_0}{a+1} \frac{\partial^2}{\partial x^2} \rho^{a+1},\qquad \mbox{with} 
 \qquad \int \rho\, \rmd x = \energy\;.
\end{equation} 
The main idea for introducing such a macroscopic description is 
the hypothesis, that the average spreading of the energy in nonlinear Hamiltonian systems of type~\eqref{eqn:general_hamiltonian} follows this NDE.
Thus, one identifies $\rho(x,t) = \langle w_k(t)\rangle$ where $k$ is understood as a discretized spatial coordinate and the averaging $\langle \cdot\rangle$ is typically taken over ensembles of trajectories and time intervals.
The essential prediction from the NDE is the one-parameter scaling of spreading with the nonlinear exponent~$a$ as the only parameter~\cite{Mulansky-Ahnert-Pikovsky-11}, which has been successfully tested by numerical studies in several cases of strongly nonlinear Hamiltonian systems~\cite{Mulansky-Ahnert-Pikovsky-11,Mulansky-Pikovsky-12,Mulansky_phd}.
The motivation for assuming a density dependent diffusion constant $D(\rho)$ in the NDE above was that the strength of chaos in the Hamiltonian lattices decreases with the energy density.
This also leads to a reduced stochasticity and thus also the diffusion constant should decrease when the energy density gets smaller.
From the purely power-law nonlinearities in the Hamiltonian system it is natural to assume a power-law dependence for the diffusion constant $D(\rho)\sim\rho^a$.

Diffusive behavior in the phase space, induced by chaos, has been studied also in low-dimensional Hamiltonian systems.
There, anomalous transport might occur due to the mixed phase space structure with regular island in a chaotic sea.
Chaotic trajectories might feel remainders of the destroyed integrability close to such regular islands which leads to so-called ``accelerator modes''~\cite{Zaslavsky-94,Shlesinger-Zaslavsky-Klafter-93}.
By analyzing the self-similarity of the structure of regular islands it was found that the diffusion process should be more precisely
described by the fractional diffusion equation (FDE)~\cite{Metzler-Klafter-00}:
\begin{equation}
 \frac{\partial^\gamma}{\partial t^\gamma} \rho = D_0 \frac{\partial^2}{\partial x^2} \rho\;,
\end{equation} 
where $\partial^\gamma/\partial t^\gamma$ denotes the fractional derivative of order $\gamma>0$ in the Caputo sense, defined later.
This fractional time derivative introduces a memory effect and thus accounts for the sticking of trajectories to surviving integrable tori in the mixed phase space.

There is no general reason why such an effect should not be seen in the strongly Hamiltonian lattices discussed here.
The phase space of coupled harmonic or nonlinear oscillators might also exhibit islands with integrable trajectories and thus possibly give rise to phenomenon describable by a fractional diffusion equation.
To account for both effects, the reduction of chaoticity due to a decreasing density and the possibly mixed phase space, we introduce here the fractional nonlinear diffusion equation (FNDE) as a phenomenological model to describe the spreading process in nonlinear Hamiltonian systems \eqref{eqn:general_hamiltonian}:
\begin{equation} \label{eqn:FNDE}
 \frac{\partial^\gamma}{\partial t^\gamma}\rho = D_0 \frac\partial{\partial x} \left(\rho^a \frac{\partial \rho}{\partial x}\right) = \frac{D_0}{a+1} \frac{\partial^2}{\partial x^2} \rho^{a+1},\qquad \mbox{with} 
 \qquad \int \rho\, \rmd x = \energy\;.
\end{equation} 
As above, $\partial^\gamma_t$ denotes the Caputo fractional derivative, defined as:
\begin{equation} \label{eqn:frac_derivative}
 \frac{\partial^\gamma \rho(x,t)}{\partial t^\gamma} = 
  \begin{cases}
    \frac{\partial^\gamma \rho(x,t)}{\partial t^\gamma} & \text{for}\quad \gamma \in\mathbb{N}\;,\\
    \frac{1}{\Gamma(n-\gamma)}\int\limits_0^t(t-\tau)^{-\gamma+n-1}\,\frac{\partial^n \rho(x,\tau)}{\partial \tau^n} \rmd \tau & \text{else}\;,
  \end{cases}
\end{equation}
with $n=\lceil\gamma\rceil\in\mathbb{N}$ being the smallest integer with  $n>\gamma$.

\subsection{Scaling properties of the FNDE}
In the following, we will analyze the FNDE to deduce its scaling predictions for spreading states.
Our analysis will closely follow previous considerations of the normal NDE~\eqref{eqn:NDE}, where the source-type solution can be found explicitly from a self-similar ansatz~\cite{Barenblatt-96}.

First, we look at the scaling properties related to a change of the conserved quantity~$\energy$.
Therefore, we assume that $\rho(x,t)$ is a solution of \eqref{eqn:FNDE}.
We rescale this solution to find a new solution $\tilde \rho(x,\tilde t)$ using a scaling parameter~$b$:
\begin{equation}
 \tilde \rho = b\rho, \qquad \tilde t = b^\alpha t,
\end{equation} 
with a scaling exponent $\alpha$ such that $\tilde \rho$ is again a solution of~\eqref{eqn:FNDE}.
A straight forward substitution of $\tilde \rho$ into the FNDE, defining $\laplace_x:=\partial_x^2$, gives the terms:
\begin{equation}
 \laplace_x \tilde\rho^{a+1} = b^{a+1}\laplace_x\rho,
\end{equation} 
and
\[
 \frac{\partial^\gamma}{\partial \tilde t^\gamma} \tilde\rho = \frac{1}{\Gamma(n-\gamma)}\int_0^{\tilde t}(\tilde t - \tilde \tau)^{-\gamma+n-1} \frac{\partial^n \tilde \rho(x,\tilde \tau)}{\partial \tilde \tau^n} \rmd \tilde \tau =b^{1-\gamma z} \frac{\partial^\gamma}{\partial t^\gamma} \rho
\]
The nonlinear fractional diffusion equation hence reads:
\begin{equation}
 b^{1-\alpha\gamma}\partial^\gamma_t \rho = b^{a+1} \frac{D_0}{a+1} \laplace_x \rho^{a+1}.
\end{equation}
This is a scaled version of \eqref{eqn:FNDE} if $ \alpha = -\frac{a}{\gamma}$.
If we now set $b=1/\energy$ we can find the scaling relation of the time that is implied when reducing a solution with arbitrary energy $\energy$ to the normalized case $\tilde \energy=1$, namely:
\begin{equation} \label{eqn:FNDE_time_energy_scaling}
 \tilde t = \energy^{a/\gamma} t\;.
\end{equation} 
Note, that this result is compatible with previous findings for the usual NDE ($\gamma=1$), where one indeed finds $\tilde t = \energy^at$.
It means that for any solution $\rho(x,t)$ with arbitrary energy $\energy$, time and energy always have to appear in the combination above~\eqref{eqn:FNDE_time_energy_scaling}.

For both, the nonlinear diffusion equation ($\gamma=1$) and the linear fractional diffusion equation ($a=0$), one finds source-type solutions by using a self-similar ansatz.
It is therefore natural to expect that this ansatz would also be successful for the FNDE~\eqref{eqn:FNDE} considered here.
Thus, we use the self-similar ansatz:
\begin{equation}
 \rho(x,t) = t^{-\mu} f(xt^{-\nu})
\end{equation} 
to identify some scaling properties of the nonlinear fractional diffusion equation.
We start with demanding the conservation of energy:
\[
 \energy = \int \rho \rmd x = \int t^{-\mu} f(xt^{-\nu}) \rmd x = t^{\nu-\mu}\int f(y) \rmd y.
\]
Hence, we conclude $\mu=\nu$ in the self-similar ansatz, because the r.h.s.\ has to be independent of time.
Considering the FNDE directly, one finds for the r.h.s.\ of~\eqref{eqn:FNDE}:
\begin{equation}
 \laplace_x \rho^{a+1} = t^{-\mu(a+1)} t^{-2\mu} \laplace_y f^{a+1}.
\end{equation} 
The fractional derivative can be evaluated as:
\[
 \partial^\gamma_t \rho = t^{-\mu} f(xt^{-\mu}) \\
  = \frac{1}{\Gamma(n-\gamma)}\int\limits_0^t(t-\tau)^{-\gamma+n-1}\,\partial^n_\tau (t^{-\mu} f(xt^{-\mu})) \rmd \tau= t^{-\gamma-\mu} y^{-1-\gamma/\mu} F(y)\;,
\]
where we use $x = yt^\mu$ and introduce the integral $F(y)$:
\begin{equation}
 F(y) = \frac{1}{\Gamma(n-\gamma)}\int\limits_0^{1/y}(y^{-1/\mu}-\tilde y^{1/\mu})^{-\gamma+n-1}\,\left(\mu\tilde y^{1-1/\mu} \right)^{n-1} \partial^n_{\tilde y} \left(\frac1{\tilde y} f(1/\tilde y)\right) \rmd \tilde y.
\end{equation}
Using these expressions, the FNDE for this self-similar ansatz gives:
\begin{equation} \label{eqn:FNDE_f}
 t^{\mu(a+2)-\gamma} y^{-1-\gamma/\mu} F(y) = \frac{D_0}{a+1} \laplace_y f^{a+1}.
\end{equation}
Thus one is left with a closed integro-differential equation for $f(y)$, if the scaling exponent is set to:
\begin{equation} \label{eqn:scaling_parameter}
 \mu=\frac\gamma{a+2}.
\end{equation} 
Here, we will not look further at solutions $f(y)$, but rather suppose that such a solution exists.
%TODO:
%\todo{cite mathematical works in this direction?}
We note that this result is consistent with self-similar solutions for the linear fractional NDE ($a=0$)~\cite{Gorenflo-Luchko-Mainardi-00}, where the scaling was found to be $\mu=\gamma/2$.
The scaling properties of such a solution then imply predictions on the spreading, namely $L\sim t^\mu$, where $L$ is some length scale of the spreading state, e.g.\ the width.
Using the result from above~\eqref{eqn:FNDE_time_energy_scaling} one can also deduce the correct energy scaling of this spreading law $L\sim(\energy^{a/\gamma}t)^\mu$, which gives the following scaling prediction for spreading:
\begin{equation} \label{eqn:FNDE_spreading}
  \frac L\energy \sim \left( \frac {t-t_0}{\energy^{2/\gamma}}\right )^{\frac{\gamma}{a+2}}
\end{equation}
Solving for $t$ and taking the derivative with respect to $L$, one also finds a scaling prediction for the excitation times:
\begin{equation} \label{eqn:FNDE_dT}
 \energy^{1-2/\gamma} \frac{\rmd t}{\rmd L} \sim \left(\frac L \energy \right)^\frac{a+2-\gamma}{\gamma}
\end{equation} 
Both results resemble the relations for the NDE with $\gamma=1$ reported earlier~\cite{Mulansky-Ahnert-Pikovsky-11,Mulansky_phd} and summarized in the next section.

\subsection{Self-similar solution of the NDE} 
\label{sec:self_sim_1d}

\begin{figure}[t]
 \begin{subfigure}[b]{0.48\textwidth}
  \includegraphics[width=\textwidth]{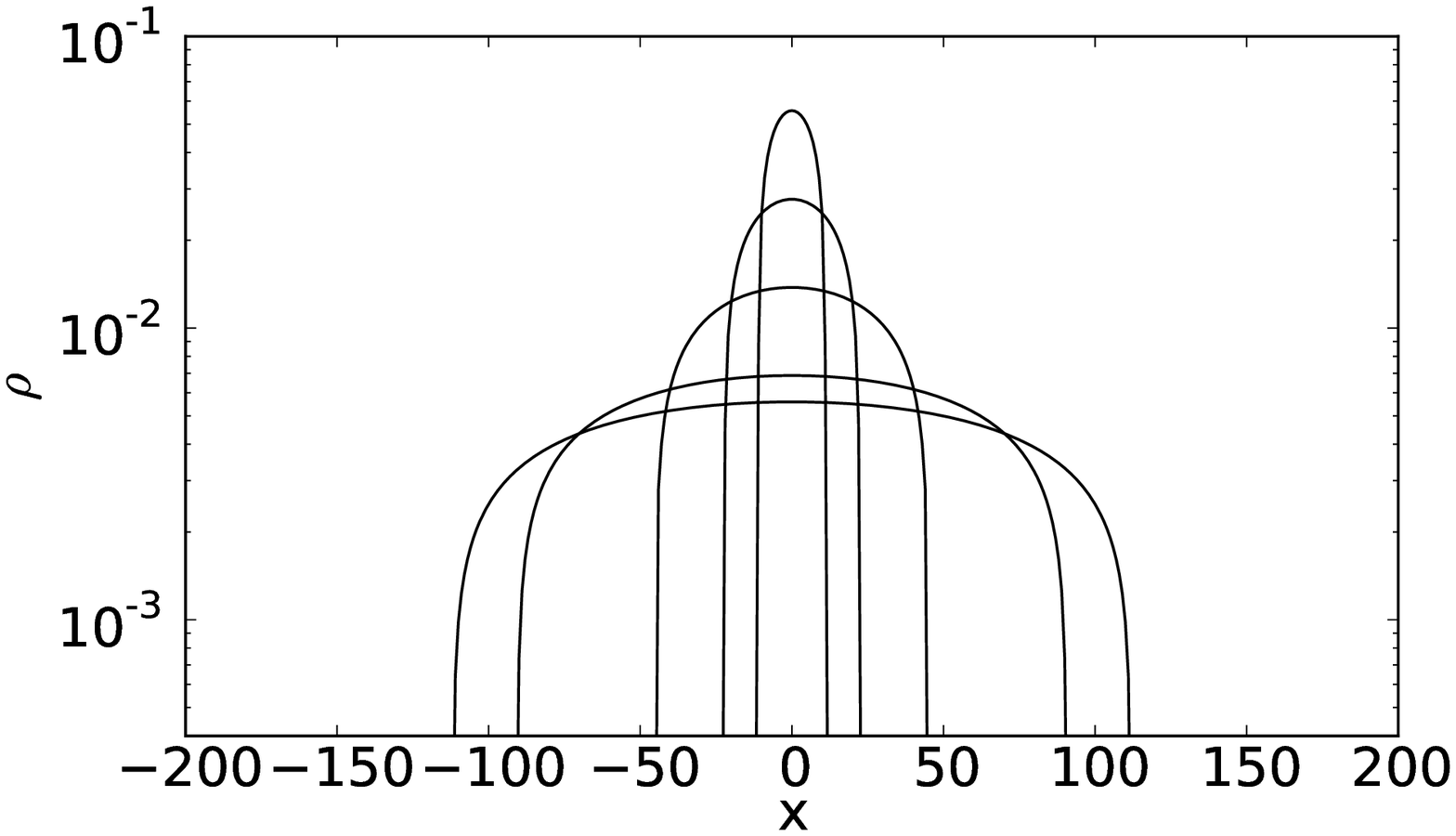}
  \vspace{0.1cm}
  \caption{(a) Self-similar solution.}
  \label{fig:self_sim}
 \end{subfigure}
 \begin{subfigure}[b]{0.48\textwidth}
  \includegraphics[width=\textwidth]{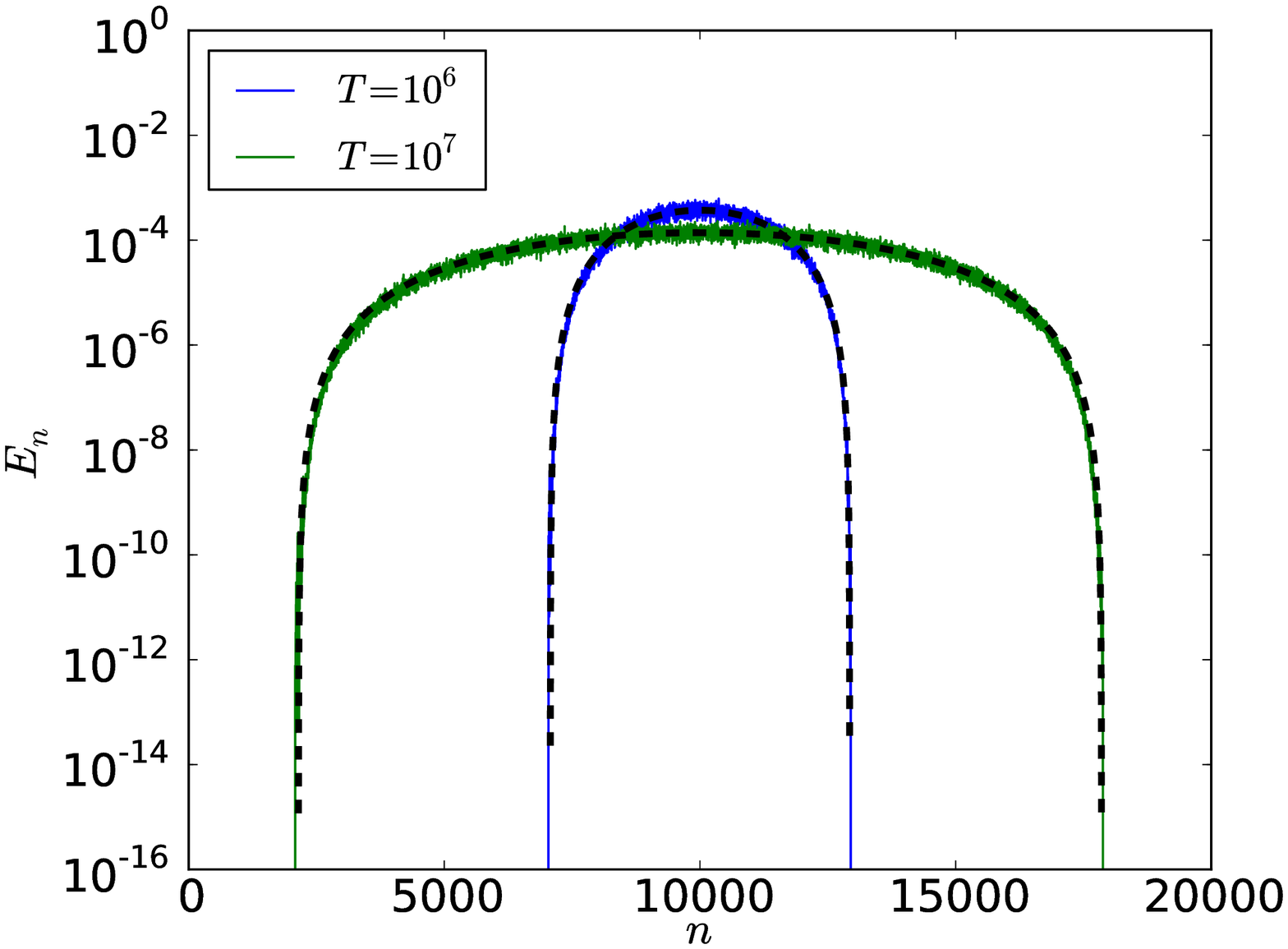} 
  \caption{(b) Numerical spreading state.}
  \label{fig:self_sim_66}
 \end{subfigure}
 \caption{(a): Self similar solution $\rho(x)$ as given by eq.~(\ref{eqn:self_sim_solution}) of the NDE for $a=2$ at times $t=10^4, 10^5, 10^6, 10^7, 10^8$ (inner to outer lines). Note the logarithmic scaling of $\rho$. (b) shows the averaged local energy excitation $\langle E_n\rangle$ of the spreading state in a nonrandom lattice at times $T=10^6$ and $T=10^7$ for $\kappa=\lambda=6$ and $\beta=4$. The average is taken over $M=48$ random initial conditions. The dashed black line shows the corresponding analytic self-similar solution of the NDE.}
 \label{fig:self_sim_1d}
\end{figure}

For the FNDE it is at the moment unclear if the profile of the self-similar solution $f(y)$ can be found analytically by solving~\eqref{eqn:FNDE_f}.
For the usual NDE where $\gamma=1$, the ordinary differential equation for the scaling function $f(y)$ is much simpler~\cite{Mulansky_phd}:
\begin{equation}
 -\mu f - \mu x\,\partial_y f = \frac{D_0}{a+1}\triangle_y f^{a+1},
\end{equation} 
with $\mu=1/(a+2)$ as above.
This ODE can indeed be solved explicitly which leads, going back to the original variables $\rho(x,t)$, to the following self-similar solution of the NDE~\cite{Polyanin-Zaitsev-03}:
\begin{equation} \label{eqn:self_sim_solution}
\rho(x,t) =
%\begin{cases} 
\left\{
\begin{array}{ll}
  (t-t_0)^{-\mu} \left(c\,\energy^{2a\mu}-\frac{a x^2}{2(a+2)(t-t_0)^{2\mu}}\right)^\frac{1}{a} & \mbox{for} \quad |x|<X(t)\;,\\
    0 & \mbox{for} \quad |x|>X(t)\;,
\end{array}
\right.
%\end{cases}
\end{equation}
with $c$ being a constant of integration which follows from the energy conservation:
\[
  c = \left( \sqrt{\frac{a}{2\pi(a+2)}}\cdot\frac{\Gamma(3/2+1/a)}{\Gamma(1+1/a)}\right)^{2a\mu}.
\]
$X(t)$ is the sharp front of the field and has the following time dependence:
\begin{equation} \label{eqn:edge_propagation}
  X(t) = \sqrt{2c\frac{2+a}a} (\energy^{a}(t-t_0))^{\frac{1}{a+2}}\;.
\end{equation}
The solution has sharp edges (see Fig.~\ref{fig:self_sim_1d}) and its spatial extension is given by $X$. 
The size of the wave packet grows in time as a power law, which can be represented in a scaling form as
\begin{equation} \label{eqn:NDE_spreading}
 \frac{X}{\energy} \sim \left(\frac{t-t_0}{\energy^2}\right)^{\frac{1}{a+2}}\;.
\end{equation}
In order to get rid of the undetermined time offset $t_0$, we
 calculate, following~\cite{Mulansky-Ahnert-Pikovsky-11}, the local inverse velocity of the spreading as $\rmd t / \rmd X$   and obtain
for it the following scaling law:
\begin{equation} \label{eqn:NDE_dT}
 \frac{1}{\energy}\frac{\mbox{d} t}{\mbox{d} X}  \sim \left(\frac{X}{\energy}\right)^{a+1}\;.
\end{equation}
Identifying the excitation edge $X$ with the spatial extent~$L$ before, one sees that this indeed corresponds with the scaling result for the FNDE above~\eqref{eqn:FNDE_spreading}, \eqref{eqn:FNDE_dT} for $\gamma=1$.

\subsection{Implications for spreading in lattices} 

In our numerical simulations of strongly nonlinear lattices, one approach is to calculate the characteristic size of the wave packet by appropriate averaging of the entropies~(\ref{eq:entrdef}). 
In particular, we can directly attribute the size of the field at a given time to the participation number, so that in (\ref{eqn:FNDE_spreading}) $L\sim \mathcal{L}$. 
Thus, if we assume that the NDE provides an adequate description of the spreading in strongly nonlinear lattices, the spreading data for different energies should fulfill scaling  (\ref{eqn:FNDE_spreading}), where the constants $\gamma$ and $a$ depend generally on the powers $\kappa,\lambda$: 
\begin{equation} \label{eqn:spreading}
 \frac{\mathcal{L}}{\energy}\sim  \left(\frac{t-t_0}{\energy^{2/\gamma}}\right)^{\frac{\gamma}{a+2}} \;.
\end{equation}

Similarly, in the second method we calculate the average time $\Delta T$ needed for spreading, in dependence of the field spatial extend $L$. This quantity directly corresponds to the inverse velocity in (\ref{eqn:FNDE_dT}): $\rmd t / \rmd L\sim\Delta T$. Thus, we expect that the times $\Delta T$ behaves as:
\begin{equation} \label{eqn:dT}
 \frac{\Delta T}{\energy^{2/\gamma-1}}  \sim \left(\frac{L}{\energy}\right)^{\frac{a+2-\gamma}{\gamma}}\;.
\end{equation}
We note that in both predictions, \eqref{eqn:spreading} and \eqref{eqn:dT}, the two influences from the fractional derivative $\gamma<1$ and the nonlinearity $a>0$ can be nicely separated.
The energy scaling in the l.h.s. of (\ref{eqn:dT}) is solely dependent on $\gamma$, so first by identifying the energy scaling in the numerical results one can determine~$\gamma$.
The power law of the subdiffusive spreading than determines the nonlinearity parameter $a$.
We already note here that in some cases we numerically find a density dependent nonlinear exponent $a(w)$ where $w$ is the energy density $w=\energy/L$~\cite{Mulansky-Ahnert-Pikovsky-11}.

\section{Results}
\label{sec:nres}
In the following sections we report on extensive numerical simulations of strongly nonlinear lattices, trying to check the predictions of the NDE framework.  For the numerical time evolution we used a 4th-order symplectic Runge-Kutta scheme \cite{McLachlan-95,Ahnert-Mulansky-11}, mostly with step-size $\Delta t = 0.1$.

\subsection{Homogeneous nonlinearity}

\subsubsection*{\bf Scaling induced spreading prediction.}

We start with the case of homogeneous nonlinearity $\kappa=\lambda$ in (\ref{eqn:general_hamiltonian}), where the local and coupling potential have the same nonlinear power:
\begin{equation}
 H = \sum_k \frac{p_k^2}{2} + \frac{\omega_k^2}{\kappa} q_k^\kappa + \frac\beta\kappa (q_{k+1} - q_k)^\kappa\;.
\end{equation} 
Here, $\beta$ is the parameter determining the relative strength between local and coupling potential and the total energy can be rescaled to unity as described in section~\ref{sec:resc} and is thus not a free parameter in the equations.
We can find a relation between the order of the fractional derivative $\gamma$, the nonlinearity of the FNDE $a$ and the parameter $\kappa$ for this homogeneous case. 
Indeed, from the scaling invariance of the Hamiltonian in (\ref{eqn:symmetric_scaling}), we obtain that the time scales with energy as:
\begin{equation}
t\sim \energy^{\frac{2-\kappa}{2\kappa}}\;.
\label{eq:hind1}
\end{equation}
On the other hand, the FNDE obeys the scaling relation: $t-t_0\sim \energy^{-a/\gamma}$~\eqref{eqn:FNDE_time_energy_scaling}.
Motivated from previous results, we assume that the FNDE gives a correct macroscopic description of the spreading process.
If this assumption is true, then the spreading states have to fulfill both scaling relations above, which gives the nonlinearity parameter $a$ as a function of $\gamma$ and $\kappa$:
\begin{equation}
 a=\gamma\,\frac{\kappa-2}{2\kappa}.
\end{equation}
To get an exact spreading prediction, one still has to obtain the parameter $\gamma$ of the fractional derivative that is introduced to account for the mixed phase space of the system.
However, in the following we will consider situations of large perturbations where it is reasonable to assume that the phase space is fully chaotic~\cite{Mulansky-Ahnert-Pikovsky-Shepelyansky-11}.
Large perturbation means that the coupling parameter is of the order $\beta\approx1$.
In this case we expect that the perturbation is strong enough to destroy all remainders of the integrability for $\beta=0$, and thus we make the assumption that $\gamma=1$.
This then gives the following spreading predictions:
\begin{equation}
 \mathcal{L}\sim (t-t_0)^{\frac{1}{a+2}}\;, \qquad \Delta T\sim L^{a+1}\;, \qquad
 a=\frac{\kappa-2}{2\kappa}\;.
\label{eq:hind2}
\end{equation}
These exact relations will serve as a test for our assumption that the NDE  adequately describes the spreading in nonlinear Hamiltonian lattices (\ref{eqn:general_hamiltonian}).

\begin{figure}[t]
 \centering
 \psfrag{xlabel1}[cc][cc]{\footnotesize $\log_{10}L$}
 \psfrag{ylabel1}[cc][cc]{\footnotesize $\log_{10}\Delta T$}
 \psfrag{xlabel3}[cc][cc]{\footnotesize $\log_{10}L$}
 \psfrag{ylabel3}[cc][cc]{\footnotesize slope $\alpha$}
 \begin{subfigure}[b]{0.48\textwidth}
  \includegraphics[width=\textwidth]{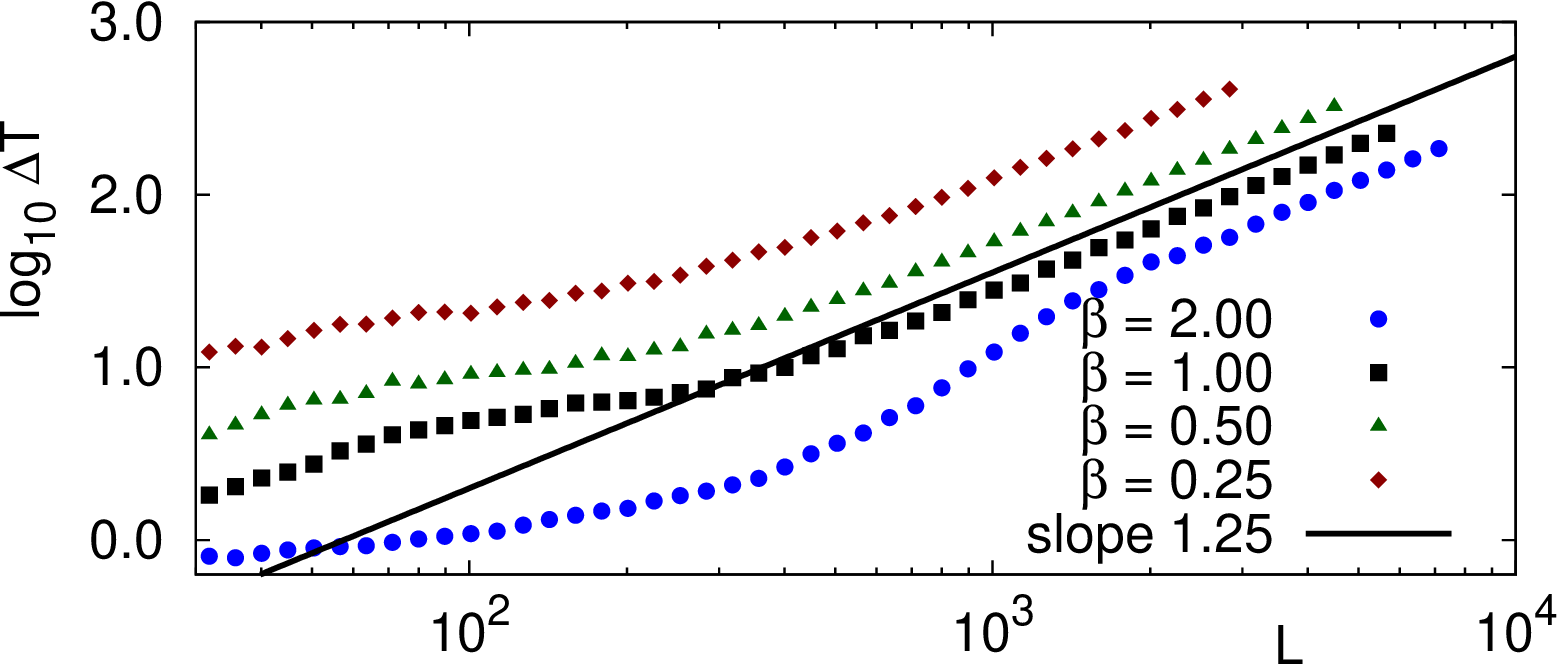}
  \caption{(a) Excitation Times for $\kappa=\lambda=4$}
  \label{fig:exc_time_44}
 \end{subfigure}
 \begin{subfigure}[b]{0.48\textwidth}
  \includegraphics[width=\textwidth]{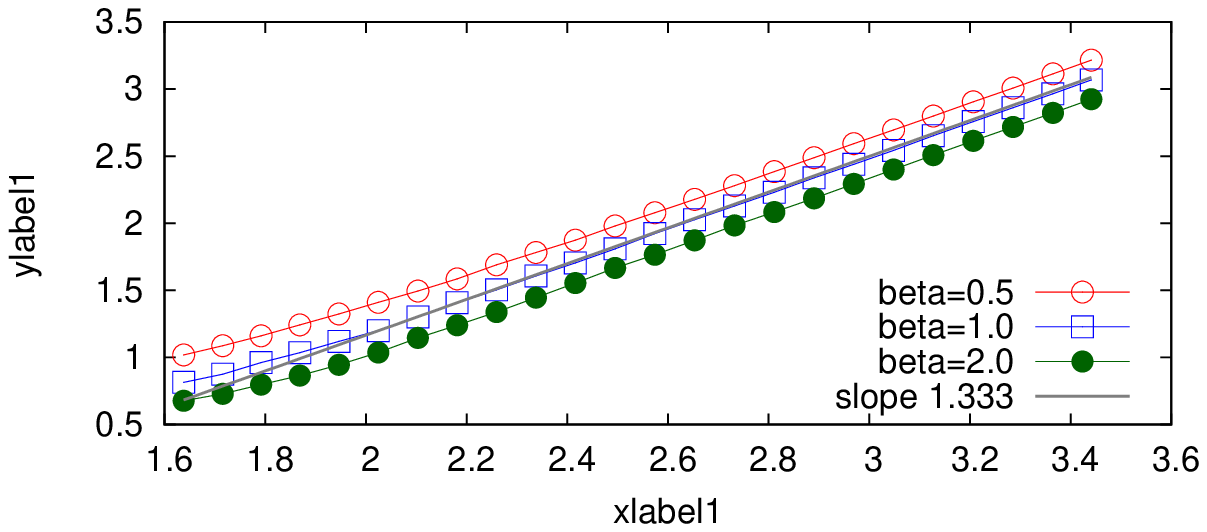}
  \caption{(b) Excitation Times for $\kappa=\lambda=6$}
  \label{fig:exc_time_66}
 \end{subfigure}
 % self_sim.eps: 0x0 pixel, -2147483648dpi, 0.00x0.00 cm, bb=
 \caption{Excitation times $\Delta T$ for the lattices with homogeneous nonlinearities, $\kappa=\lambda=4$ (left panels) and $\kappa=\lambda=6$ (right panels). 
 The black dashed lines in the upper graphs show the predicted behaviors [see (\ref{eq:hind2})] $\Delta T \sim L^\nu=L^{5/4}$ (left) and $\Delta T \sim L^\nu L^{4/3}$ (right) respectively.
%% The lower panels show instantaneous exponents $\alpha$ from $\Delta T \sim %L^\alpha$ approximated from finite differences of the numerical results(?).
 }
 \label{fig:exc_time_44_66}
\end{figure}

\subsubsection*{\bf Comparison with numerical results.}
To test the theoretical predictions (\ref{eq:hind2}) we follow the evolution in a one-dimensional lattice with $\omega_k \in [0,1]$, started from a single site (or several sites for $\kappa=6$) excitation in the middle.
For several values for the  nonlinear strength~$\beta = 0.25,0.5,1,2$ we integrated the system up to $T=10^6$ and analyzed the spreading in terms of $P(t)$ and $\Delta T(L)$.
This was repeated for $M=100$ realizations of disorder.
Fig.~\ref{fig:exc_time_44_66} shows the averaged results of these runs for the excitation time~$\Delta T(L)$ for $\kappa=4$ (left panels) and $\kappa=6$ (right panels).
In both cases we find a quite nice agreement of the numerical results with the analytic predictions of the NDE (\ref{eq:hind2}).
We also performed simulations choosing the disorder to be $\omega_k \in [0.5,1.5]$ and got similar results (not presented here).
Additionally, results for the direct spreading measure $P(t)$ were obtained which show the same agreement with prediction (\ref{eq:hind2})
and are also omitted here~\cite{Mulansky-Ahnert-Pikovsky-11}.
To our opinion, the agreement between numerics and prediction is a rather convincing evidence that the NDE provides the proper framework to model the average energy spreading in nonlinear Hamiltonian chains.

\begin{figure}[t]
 \begin{subfigure}[b]{0.48\textwidth}
  \includegraphics[width=\linewidth]{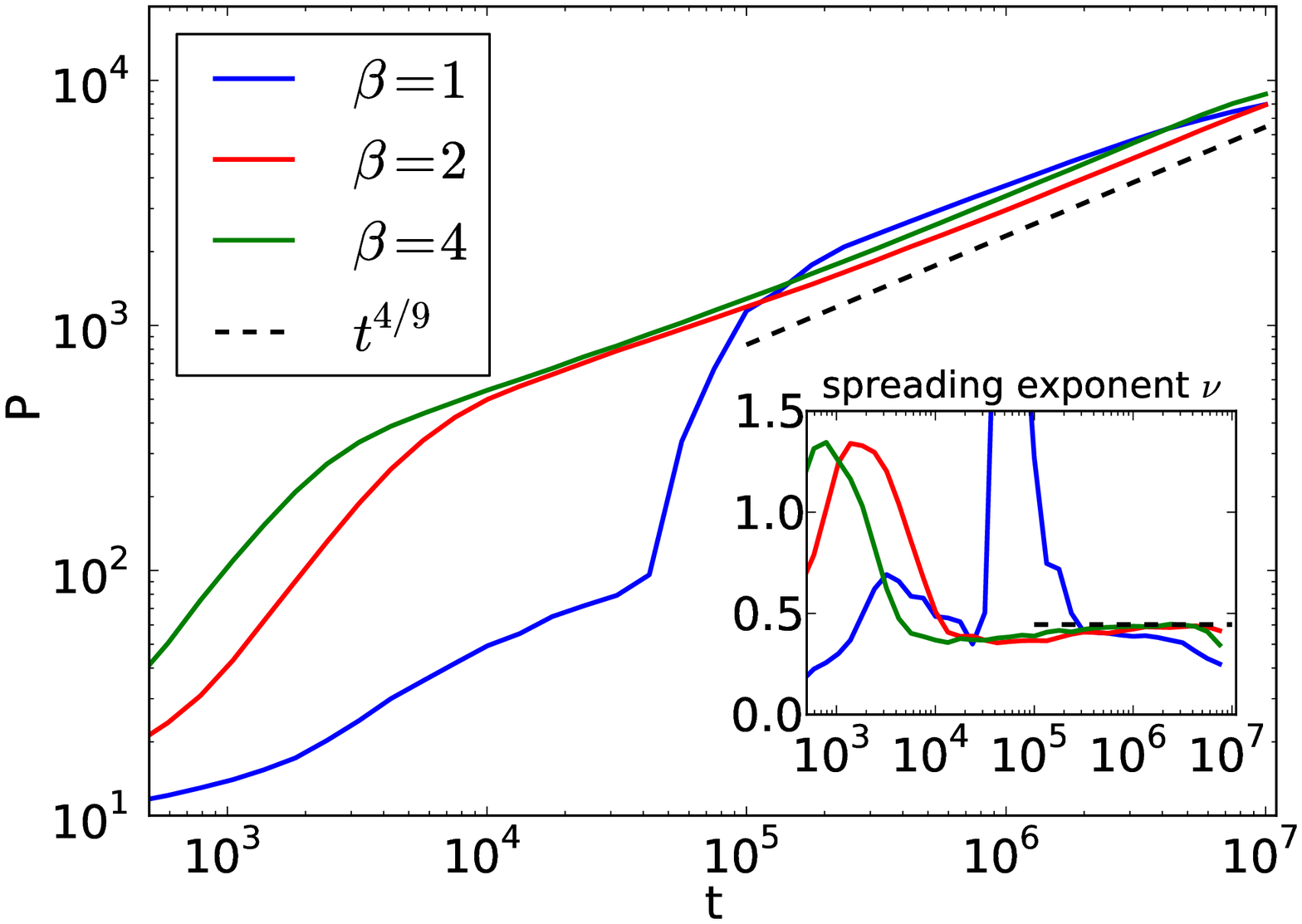}
  \caption{(a) Regular lattice with $\kappa=\lambda=4$.} \label{fig:snol_part_44}
 \end{subfigure} \hfill
 \begin{subfigure}[b]{0.48\textwidth}
  \includegraphics[width=\linewidth]{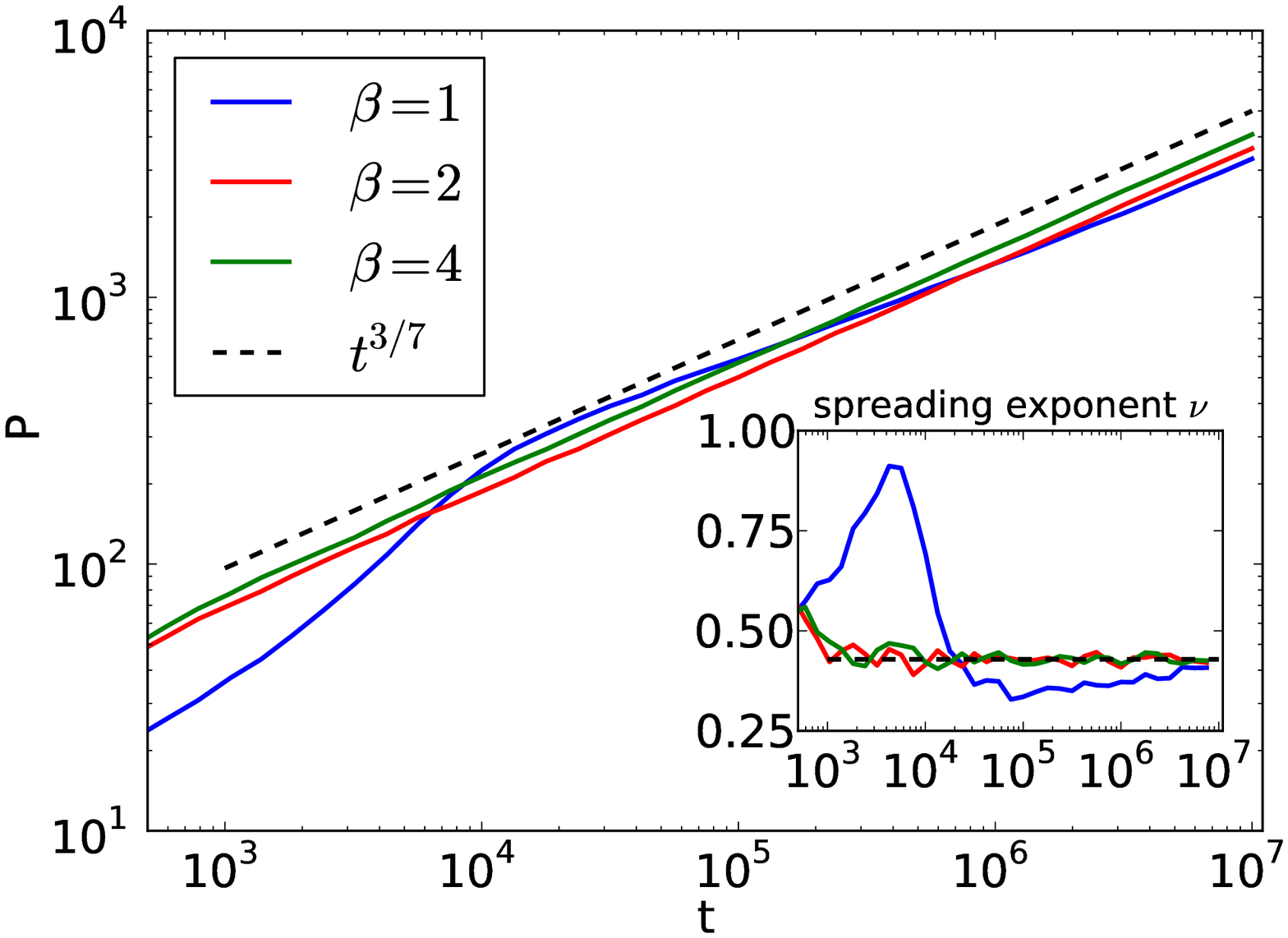}
  \caption{(b) Regular lattice with $\kappa=\lambda=6$.}
 \end{subfigure}
 % part44.eps: 0x0 pixel, 300dpi, 0.00x0.00 cm, bb=
 \caption{Participation number for a nonrandom lattice with $\kappa=\lambda=4$ (a) and $\kappa=\lambda=6$ (b) averaged over $M=48$ random initial conditions (solid lines). The dashed lines shows the NDE prediction $P\sim t^{4/9}$ and $P\sim t^{3/7}$ respectively. In the inset we plot the numerical spreading exponent $\nu$ obtained from finite differences of the method above, the dashed line there also corresponds to the expectation from the NDE.}
 \label{fig:snol_part}
\end{figure}

Remarkably, the NDE scaling  (\ref{eq:hind2}) holds also for one-dimensional lattices without disorder.
Fig.~\ref{fig:snol_part} shows the participation number evolution for
lattice (\ref{eqn:ghrk})  with $\omega_k = 1$ and $\kappa=4,6$. In this case the
excitation times $\Delta T$ are not the proper measures as they are dominated by quasi-compactons (cf. Fig.~\ref{fig:4_6_wf}(a)), but the participation number calculations are insensitive to such modes. For them we 
expect from (\ref{eq:hind2})  the scaling $P\sim (t-t_0)^{\frac{2\kappa}{5\kappa-2}}$ which is confirmed in Fig.~\ref{fig:snol_part}.
This is also supported by the direct comparison of spreading states from the numerical simulation to the self similar solution, as can be seen in right panel of Fig.~\ref{fig:self_sim_1d}.
Note that in the end of the simulation for $\beta=1,2$ in Fig.~\ref{fig:snol_part_44}, the spreading state has hit the lattice boundaries leading to a saturation of the participation number and a decrease of the spreading exponent.

Summarizing these results, we have found that from the assumption of the validity of the NDE we derived an exact spreading predictions for a fully chaotic phase space in Hamiltonian lattices with homogeneous nonlinearity $\kappa=\lambda$.
These predictions were to a high accuracy verified as the asymptotic behavior in numerous numerical simulations.
We note that this spreading process can also be observed in the case of a regular on-site potential were disorder is completely absent, Fig.~\ref{fig:snol_part}.
This shows that disorder is not required for the spreading phenomenon, an observation already made for 2D lattices in~\cite{Mulansky-Pikovsky-12}.
Hence the subdiffusive spreading is not a result of the interplay between nonlinearity and disorder, but rather a more general phenomenon lately called ``Chaotic Diffusion''~\cite{Mulansky-Pikovsky-12,Mulansky_phd}.
To further verify the assumption of $\gamma=1$ above due to the fully chaotic phase space it would be very interesting to study the behavior for smaller $\beta$.
If our argument is correct one would expect some dependence $\gamma(\beta)$ where $\gamma$ also decreases for smaller values of $\beta$.
This will be the subject of future studies.

\subsection{Nonlinear Oscillator, Nonlinear Coupling }

\subsubsection*{\bf Numerical Results.}
After having found that the NDE provides a good description of the spreading for the case of homogeneous nonlinearity, we turn now to a general situation with $\kappa \neq \lambda$.
In this section we study the case of fully nonlinear oscillators, hence we choose $\kappa=4$ and $\lambda=6,8$.
In this case, the disorder parameter $\omega_k$ in (\ref{eqn:ghr}) does not have the meaning of an oscillator frequency, but is the coefficient determining the nonlinear strength. 
The real frequency of oscillations depends on the local energy at the site. 
In Fig.~\ref{fig:4_6_wf} we show an exemplary time evolution of an initially localized state in such a lattice.
For this non-homogeneous case, the energy~$\energy$ is the crucial parameter in the Hamiltonian.
That allows us to independently determine the parameter $\gamma$ and $a$ from numerical simulations by first identifying the energy-scaling of the spreading and then computing the slope of the subdiffusive process.
That means we will compare the numerical results with the spreading predictions from \eqref{eqn:spreading} and \eqref{eqn:dT}.

\begin{figure}[tb]
 \centering
 \psfrag{xlabel1}[cc][cc]{\footnotesize $\log_{10}L$}
 \psfrag{ylabel1}[cc][cc]{\footnotesize $\log_{10}\Delta T$}
 \psfrag{xlabel4}[cc][cc]{\footnotesize $\log_{10}L/\energy$}
 \psfrag{ylabel4}[cc][cc]{\footnotesize $\log_{10}\Delta T/\energy^{0.85}$}
 \begin{subfigure}[b]{0.48\textwidth}
  \includegraphics[width=\textwidth]{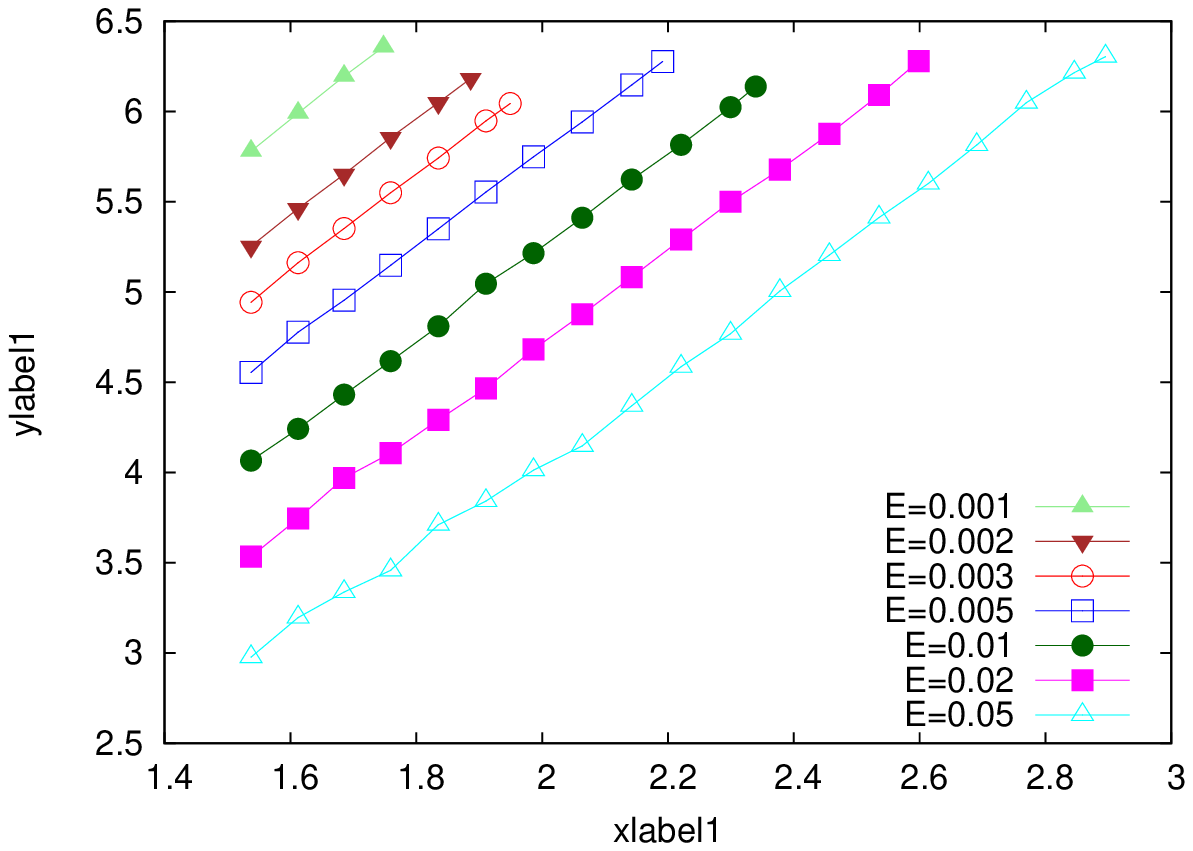}
  \caption{(a) Unscaled results for $\kappa=4$, $\lambda=6$.}
  \label{fig:4_6_unscaled}
 \end{subfigure}
 \begin{subfigure}[b]{0.48\textwidth}
  \includegraphics[width=\textwidth]{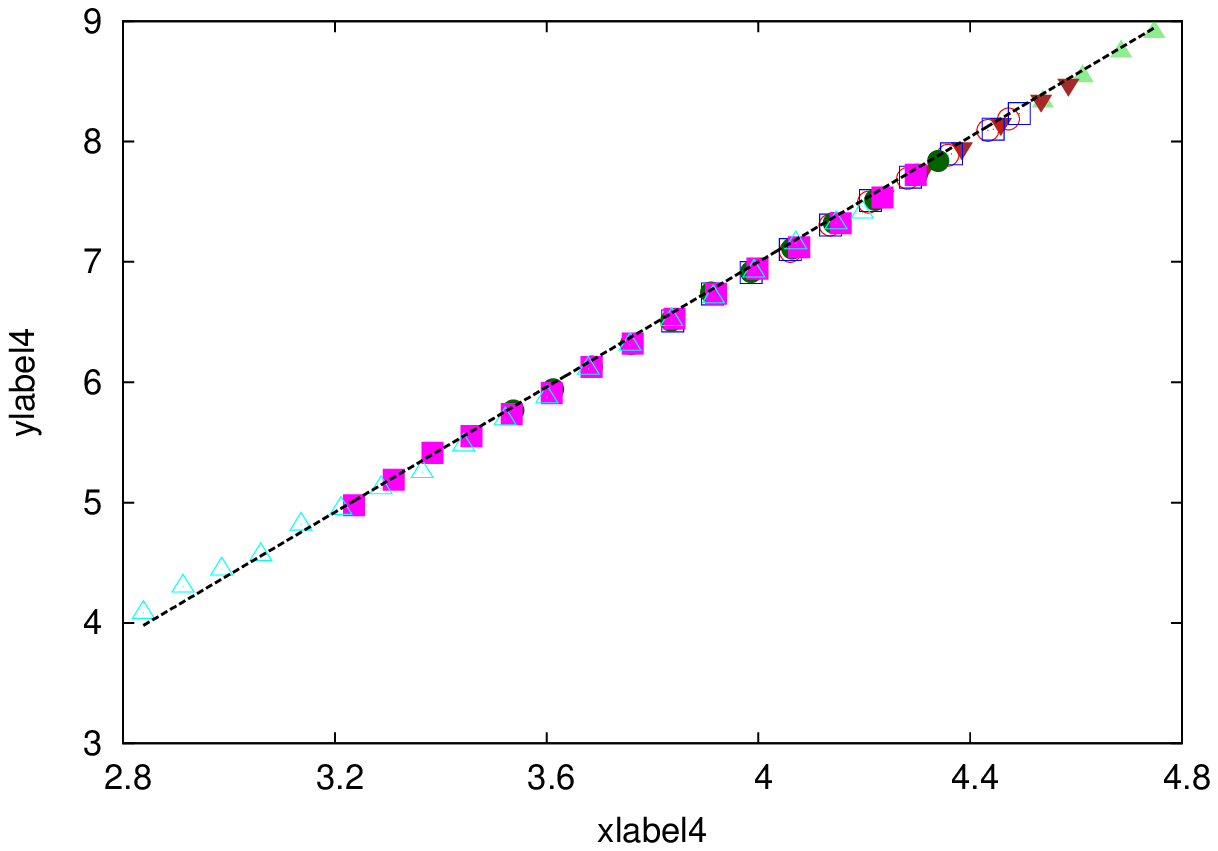}
  \caption{(b) Scaled results for $\kappa=4$, $\lambda=6$.}
  \label{fig:4_6_scaled}
 \end{subfigure}
 % self_sim.eps: 0x0 pixel, -2147483648dpi, 0.00x0.00 cm, bb=
 \caption{Excitation times $\Delta T(L)$ for $\kappa=4$, $\lambda=6$ with energies $\energy=0.001\dots0.05$. Panel (a) shows the original data while in panel (b) you see scaled variables as suggested by the FNDE with $\gamma=1.08$: $\Delta T/\energy^{0.85}$ vs.\ $L/\energy$. The dashed line has slope $(a+2-\gamma)/\gamma=2.6$.
 }
 \label{fig:4_6}
\end{figure}

We start with the case $\kappa=4$ and $\lambda=6$ and investigate the excitation times $\Delta T(L)$ for different total energies.
The results of our simulations for $\omega_k\in [0.5,1.5]$ are shown in Fig.~\ref{fig:4_6}.
In the right panel the scaling as suggested by the FNDE \eqref{eqn:dT} is applied and we found the best overlap of the individual curves for the parameter value $\gamma=1.08$ that gives the scaling exponent $1-2/\gamma\approx0.85$.
That means we find only a slight deviation from the pure NDE case where $\gamma=1$, hence the influence of the mixed phase space is rather small, but clearly identifiable as for $\gamma=1$ the curves do not overlap as perfectly (cf.~\cite{Mulansky-Ahnert-Pikovsky-11}).
The numerical data also nicely follow a straight line as seen in Fig.~\ref{fig:4_6_scaled}.
This indicates subdiffusive behavior with a slope $(a+2-\beta)/\beta\approx2.6$ from a numerical fit and we thus calculate the nonlinear exponent in the FNDE as $a\approx1.8$.

\begin{figure}[tb]
 \centering
 \psfrag{xlabel1}[cc][cc]{\footnotesize $\log_{10}L/\energy$}
 \psfrag{ylabel1}[cc][cc]{\footnotesize $\log_{10}\Delta T/\energy$}
 \psfrag{xlabel2}[cc][cc]{\footnotesize $\log_{10}L/\energy$}
 \psfrag{ylabel2}[cc][cc]{\footnotesize $\log_{10}\Delta T/\energy^{0.7}$}
 \psfrag{xlabel3}[cc][cc]{\footnotesize $-\log_{10}w$}
 \psfrag{ylabel3}[cc][cc]{\footnotesize $\frac{a+2-\gamma}{\gamma}$}
 \psfrag{title1}[cc][cc]{\footnotesize }
 \begin{subfigure}[b]{0.48\textwidth}
  \includegraphics[width=\textwidth]{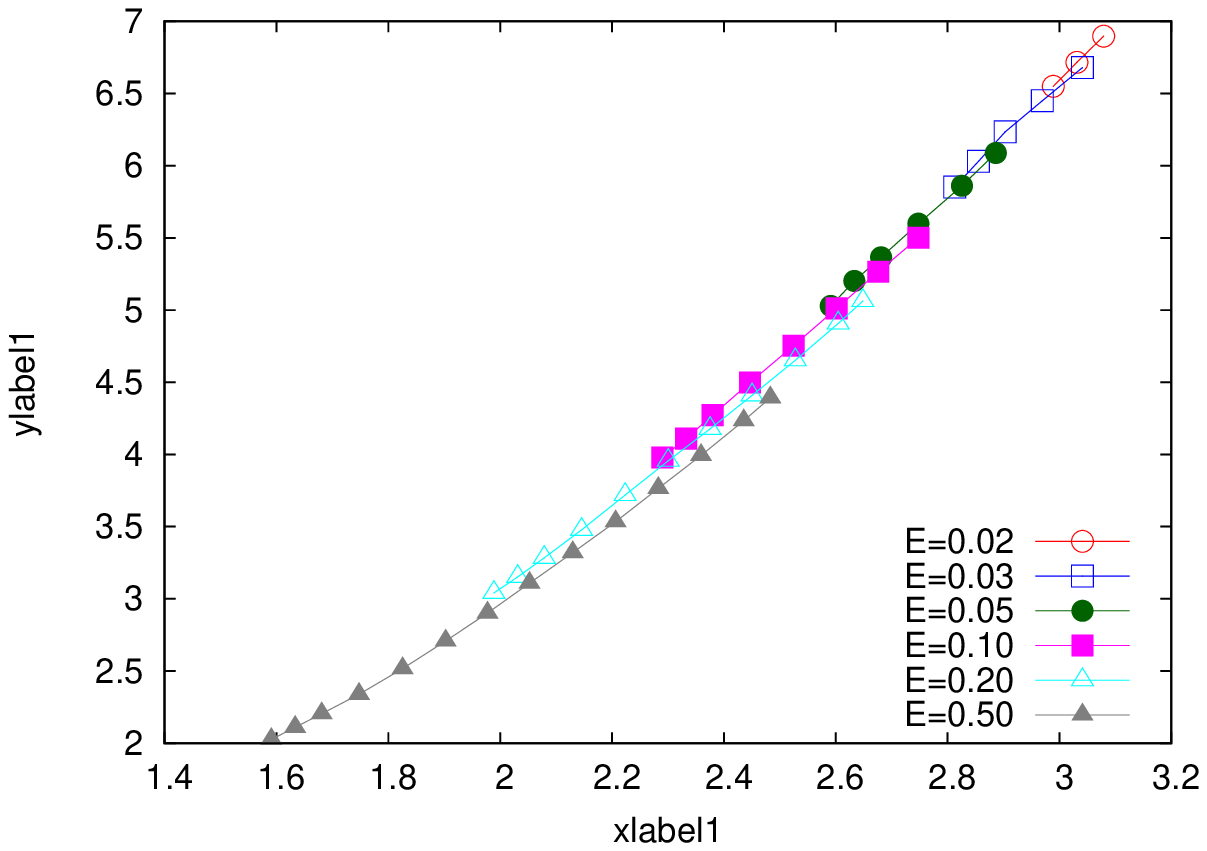}
  \caption{(a) $\kappa=4$, $\lambda=8$ scaled with $\gamma=1$}
  \label{fig:4_8_scaled_2}
 \end{subfigure}
 \begin{subfigure}[b]{0.48\textwidth}
  \includegraphics[width=\textwidth]{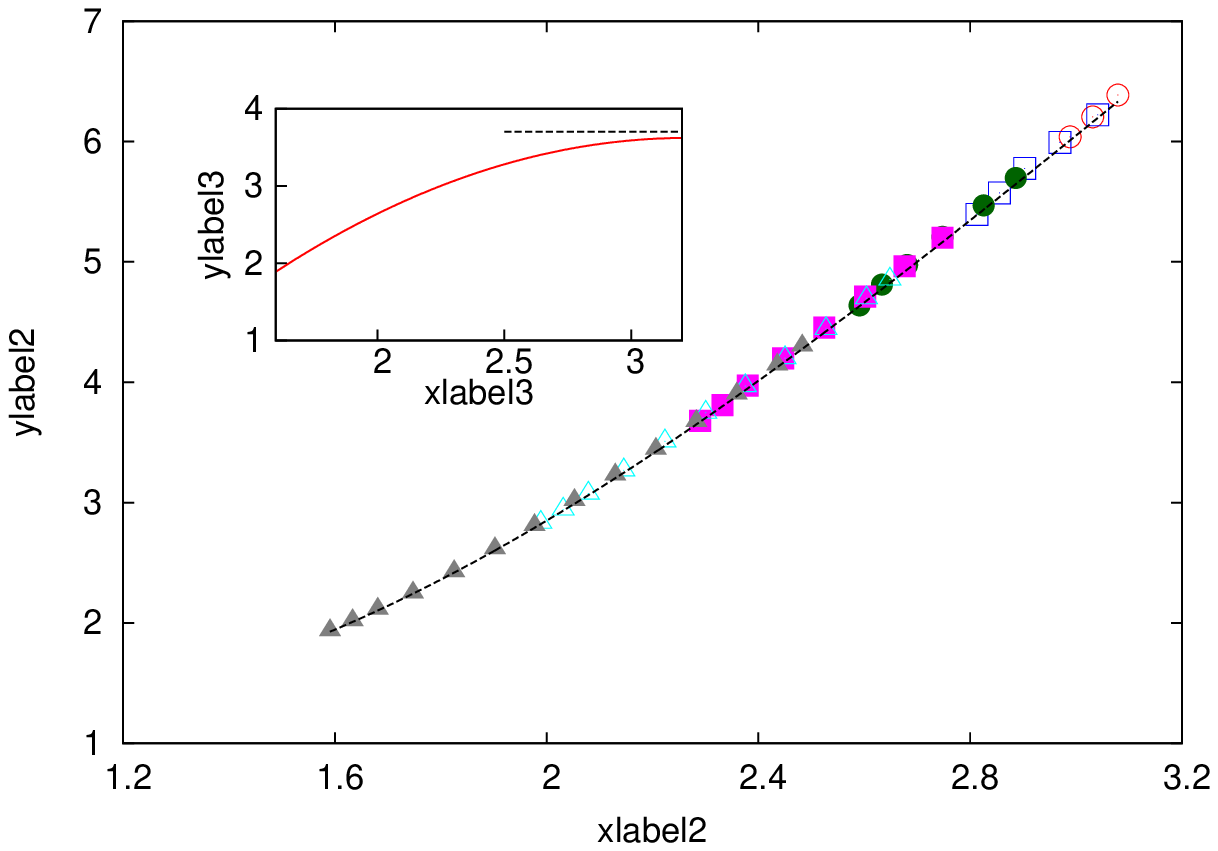}
  \caption{(b) $\kappa=4$, $\lambda=8$ scaled with $\gamma=1.18$}
  \label{fig:4_8_scaled}
 \end{subfigure}
 % self_sim.eps: 0x0 pixel, -2147483648dpi, 0.00x0.00 cm, bb=
 \caption{Excitation times $\Delta T(L)$ for $\kappa=4$, $\lambda=8$ with energies $\energy=0.02\dots0.5$. Panels (a) and (b) show the scaling of the FNDE: $\Delta T/\energy^{2/\gamma-1}$ vs.\ $L/\energy$ for two parameter values: $\gamma=1$ (a) and $\gamma=1.18$ (b). The inset in (b) shows the slope of the scaled data from a polynomial fit. The straight dashed line in this inset represents the asymptotic prediction $(a+2-\gamma)/\gamma\approx 3.7$ from microscopic dynamics.
 }
 \label{fig:4_8}
\end{figure}

In a second simulation we studied the case $\kappa=4$ and $\lambda=8$.
The results are shown in Fig.~\ref{fig:4_8} where two scalings with $\gamma=1$ (Fig.~\ref{fig:4_8_scaled_2}) and $\gamma=1.18$ (Fig.~\ref{fig:4_8_scaled}) are compared.
It is clear from these graphs that the normal NDE with $\gamma=1$ does not predict the correct scaling as the curves for different energies do not overlap in Figure~\ref{fig:4_8_scaled_2}.
For $\gamma=1.18$, the scaled variables according to the FNDE are $\Delta T/\energy^{0.7}$ vs.\ $L/\energy$ and Fig.~\ref{fig:4_8_scaled} shows that for this choice indeed a convincing overlap of the individual curves is observed.
In contrast to the case $\lambda=6$, the scaled curve for $\lambda=8$ does not follow a straight line.
We explain this by the fact that we have not reached the asymptotic regime yet in this study.
Indeed, the numerically accessible parameter range for the energy density $\energy/L$ goes only down to $\energy/L\approx10^{-3}$ in Figure~\ref{fig:4_8_scaled}, while for $\lambda=6$ we were able to go almost two orders of magnitudes lower.
To still quantify the slope in this case we performed a polynomial fit of the scaled data and plotted the derivative of this fitted curve in the inset in Figure~\ref{fig:4_8_scaled}.
The result indicates a convergence of this slope and hence an asymptotically constant value for $a\approx 3.5$.
The dashed line in this inset represents our theoretical prediction for this asymptotic value to be explained in the next subsection.

In summary, we have found here that for fully nonlinear oscillators, $\kappa=4$ and $\lambda=6,8$, the spreading of initially localized excitations can be nicely described by the FNDE.
With the scaling approach we were able to separate the two parameters $\gamma$ and $a$ of the FNDE and determine their values from the numerical results on the excitation times.
The power of the fractional derivative was obtained as $\gamma=1.08$ for $\lambda=6$ and $\gamma=1.18$ for $\lambda=6$.
For $\lambda=6$ the scaled spreading was found to behave as a power law with some slope $(a+2-\gamma)/\gamma\approx2.6$ which gives the nonlinearity parameter of the FNDE as $a\approx1.8$.
For $\lambda=8$ we could not reach the asymptotic behavior and hence found a density dependent slope, but a numerical estimation of this slope indicates for a convergence against the value $(a+2-\gamma)/\gamma\approx3.7$ which means $a\approx 3.5$.
We conclude that the FNDE is a good model to describe spreading in fully nonlinear Hamiltonian systems.

\subsubsection*{\bf Microscopic spreading dynamics.}
In \cite{Mulansky-Pikovsky-12} a microscopic model of spreading was developed that lead to an exact prediction of the spreading exponent for a regular two-dimensional lattice ($\omega_k=1$) of harmonic oscillators with nonlinear coupling.
Here, we will follow this idea and try to find a reduced system that describes the dynamics at the excitation edge.
The idea is to understand the mechanism of how a new oscillator is excited from the chaotic forcing induced by its already excited neighbor.

For harmonic oscillators with $\omega_k=1$, the situation was particularly easy because all oscillators were in resonance due to the absence of disorder.
Therefore, in~\cite{Mulansky-Pikovsky-12} it was enough to consider only two coupled oscillators at the edge, one excited and one at rest, to obtain a correct spreading prediction.
For the nonlinear oscillators with $\kappa=4$ studied here it is immediately clear that considering only two oscillators will not be sufficient.
The Hamiltonian for two coupled oscillators is:
\begin{equation} \label{eqn:two_osc_H46}
 H = \frac{p_1^2 + p_r^2}2 + \frac{q_1^4+q_r^4}{4} + \frac1\lambda(q_1 - q_r )^\lambda,
\end{equation} 
where $q_1,p_1$ denote the already excited oscillator with a local energy density $w\approx p_1^2/2+q_1^4/4$ while the second oscillator is at rest: $q_r=p_r=0$ with a zero energy density $w_r=0$.
Because the second oscillator is subject to a non-resonant forcing it will, for small energy densities $w$, only become excited up to an energy density according to standard perturbation expansion which means $w_r \sim w^{\lambda/4} \ll w$.
Hence, for small densities there is almost no energy transport from the excited to the resting oscillator which would imply that spreading should stop because no new oscillators get excited.
%\todo{better explanation?}
This prediction is clearly wrong as is seen from numerical spreading results presented above.
The reason is that the two oscillator model is too simple to describe the spreading process.
Thus, we consider more complex situations with $N$ oscillators, were the first $N-1$ oscillators are excited with some energy density $w$, while the last oscillator is at rest.
From examining the geometric properties of the resonances of such coupled nonlinear oscillators it can be argued that only for $N\geq5$ energy transport to the last oscillator through a global chaotic layer is expected~\cite{Mulansky_phd}.
%TODO:
%\todo{More details on that?}

\begin{figure}[tb]
 \centering
 \begin{subfigure}[b]{0.48\textwidth}
  \includegraphics[width=0.8\textwidth]{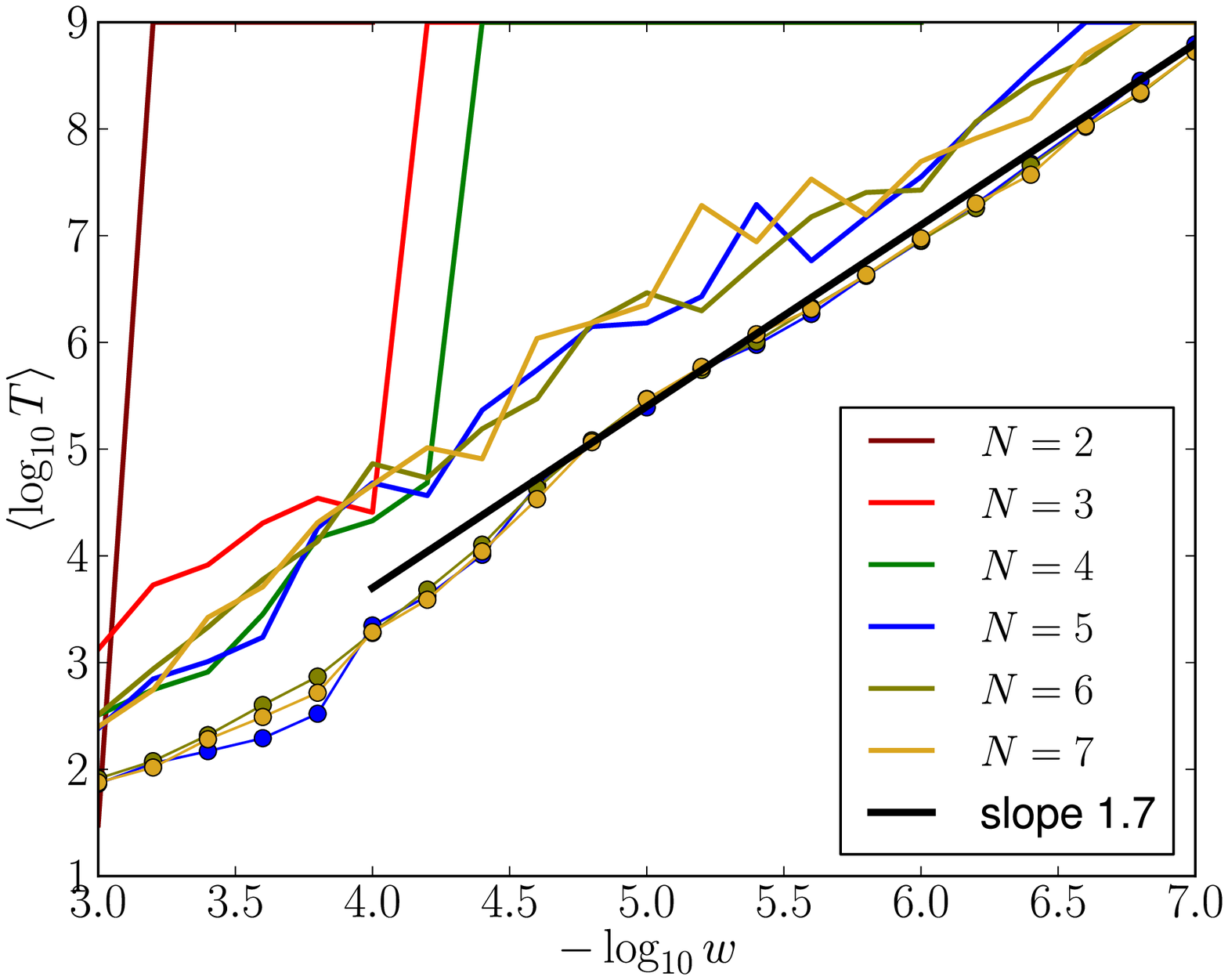}
  \caption{(a) Microscopic excitation for $\kappa=4$, $\lambda=6$}
  \label{fig:few_osc_4_6}
 \end{subfigure}
 \begin{subfigure}[b]{0.48\textwidth}
  \includegraphics[width=0.8\textwidth]{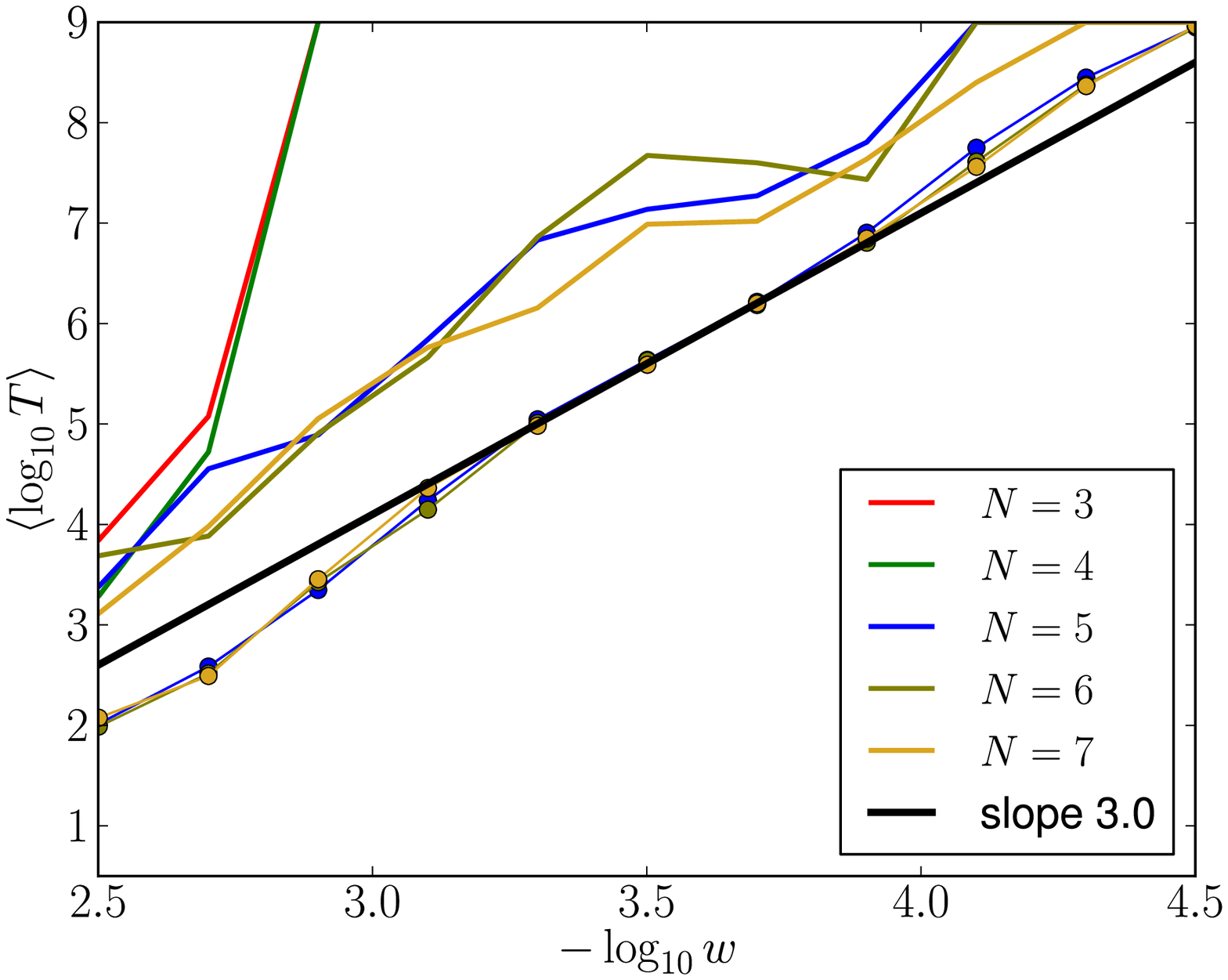}
  \caption{(b) Microscopic excitation for $\kappa=4$, $\lambda=8$}
  \label{fig:few_osc_4_8}
 \end{subfigure}
 % self_sim.eps: 0x0 pixel, -2147483648dpi, 0.00x0.00 cm, bb=
 \caption{Excitation times as a function of energy density $T(w)$ for the microscopic model of $N$ oscillators. Solid lines represent maximum values of the Monte-Carlo ensemble study and circles correspond to logarithmic ensemble averages $\langle \log_{10} T\rangle$. 
 }
 \label{fig:few_osc}
\end{figure}
Here, we will verify this conjecture by a numerical simulation.
Consider a situation with $N-1$ excited oscillators with an energy density $w$ as described above.
One way to quantify the energy transport to the last, resting oscillator is by measuring the time $T$ that is required for this oscillator to become excited to some critical energy density above the perturbative description.
This time is very similar to the excitation times introduced earlier to quantify spreading.
Here, we will fix the number of excited oscillators and just measure the time as a function of the energy density $T(w)$.
If $T$ diverges then no energy transfer beyond the perturbative excitation is taking place.
In Figure~\ref{fig:few_osc} we show the results from a Monte-Carlo study on $T(w)$ for an ensemble of $M=100$ random initial conditions and different numbers of oscillators $N=2\dots7$.
The bold lines correspond to the maximum times $\max\, T$ from this ensemble of initial conditions for each $N$ and density $w$.
In both cases, $\lambda=6$ (Figure~\ref{fig:few_osc_4_6}) and $\lambda=8$ (Figure~\ref{fig:few_osc_4_8}), one definitely observes a divergence of $T$ for $N<5$.
For $N\geq 5$, however, we found an asymptotic power-law dependence $T(w)\sim w^\chi$ with $\chi \approx -1.7$ for $\lambda=6$ and $\chi\approx -3.0$ for $\lambda=8$.
So firstly we note that the microscopic model with $N\geq5$ predicts spreading with an asymptotic power-law behavior.
To connect these numerical results from the microscopic dynamics to the macroscopic spreading one can identify the microscopic and the macroscopic excitation times $\Delta T\sim T$.
Noting that the number of oscillators in the microscopic remains constant and only the energy density changes one finds the following prediction for the macroscopic excitation time $\Delta T \sim \energy^\chi$.
Translating this into the scaled variables used earlier one finds that $(a+2-\gamma)/\gamma=2/\gamma-1-\chi$ and thus $a/\gamma=-\chi$.
For $\lambda=6$ the nonlinear exponent was calculated from the numerical spreading as $a/\gamma\approx 1.7$, which is in a very good agreement with the microscopic result $-\chi \approx 1.7$ shown in Figure~\ref{fig:4_6_scaled}.
For $\lambda=8$ the asymptotic behavior of the macroscopic spreading is also in very good agreement with these microscopic results as $-\chi\approx 3$ appears to be very close to the asymptotic sprading behavior where $a/\gamma\approx3$ in Figure~\ref{fig:4_8_scaled}.
Thus we conclude that a microscopic model of $N=5$ oscillators is enough to understand the macroscopic spreading properties in long, macroscopic chains of such oscillators.
However, at this point the exponent $\chi$ was only obtained from numerical results and analytical treatments remain a challenge for future work.

\subsection{Harmonic Oscillators, Nonlinear Coupling}

Finally, we turn to the most complicated situation of harmonic oscillators with random frequencies and nonlinear coupling.
Therefore, we assume the on-site potential to be quadratic $\kappa=2$, for the coupling we chose $\lambda = 4$ and $\lambda = 6$. 
This case corresponds to a rather general situation of nonlinear disordered lattices, where in the representation of linear eigenmodes one can also interpret the system as an ensemble of nonlinearly coupled linear modes.
The most prominent example of such a situation is the Discrete Anderson Nonlinear Schr\"odinger Equation (DANSE-model)~\cite{Pikovsky-Shepelyansky-08} which, if treated in the eigenmode basis, consists of localized
harmonic modes with nonlinear coupling.
The main difference between this setup and the strongly nonlinear lattices considered here is that in the DANSE-model the coupling between the modes has random coefficients and is exponentially decaying in space due to the overlap integrals between the localized modes.
In contrast, the strongly nonlinear lattices studied here have only a nearest neighbor coupling without a random coupling coefficient.
Similar to the studies before, we analyze the excitation times $\Delta T(L)$ as function of excitation length $L$ for different energies $\energy$ to check the predictions of the FNDE scaling (\ref{eqn:FNDE_dT}).

\begin{figure}[tb]
 \centering
 \psfrag{xlabel1}[cc][cc]{\footnotesize $\log_{10}L$}
 \psfrag{ylabel1}[cc][cc]{\footnotesize $\log_{10}\Delta T$}
 \psfrag{xlabel2}[cc][cc]{\footnotesize $\log_{10}L/\energy$}
 \psfrag{ylabel2}[cc][cc]{\footnotesize $\log_{10}\Delta T/\energy$}
 \psfrag{xlabel3}[cc][cc]{\footnotesize $-\log_{10}w$}
 \psfrag{ylabel3}[cc][cc]{\footnotesize $a(w)$}
 \begin{subfigure}[b]{0.48\textwidth}
  \includegraphics[width=\textwidth]{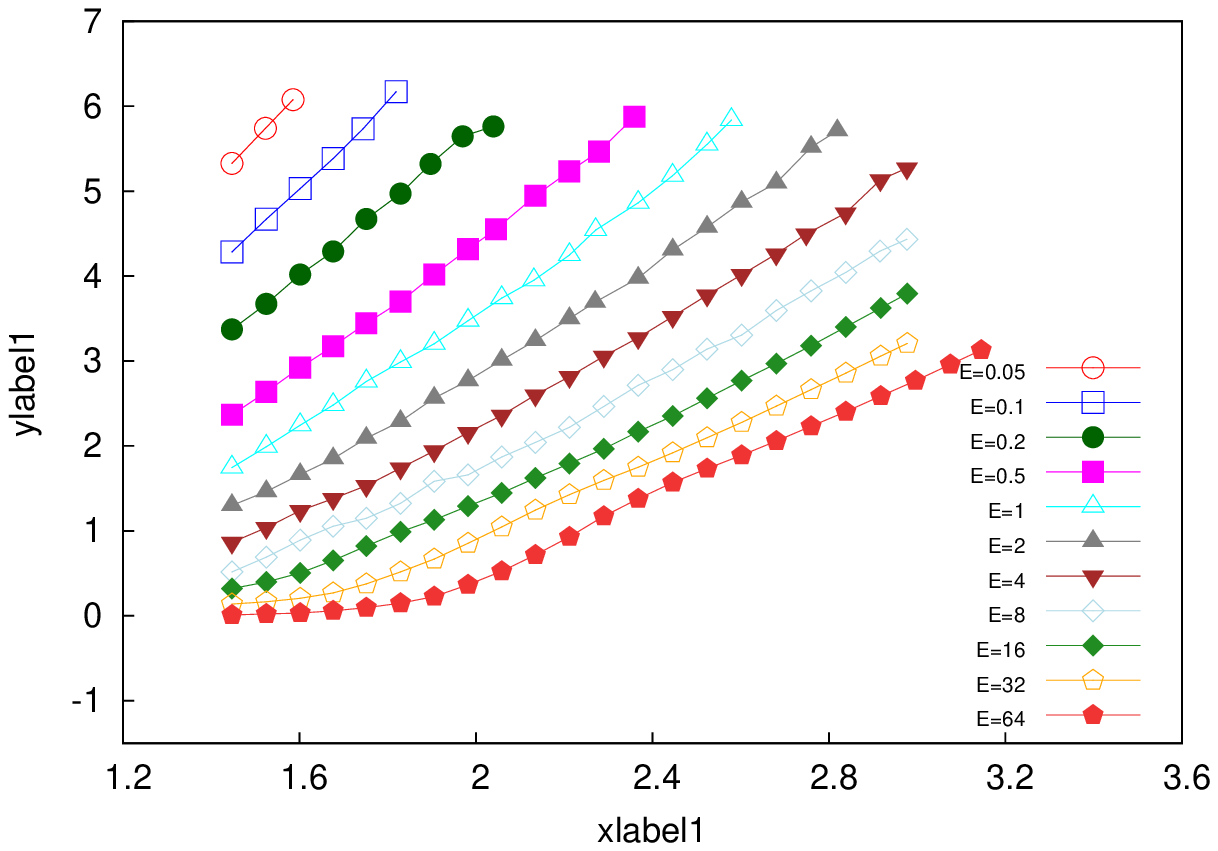}
  \caption{(a) $\Delta T(L)$ for $\kappa=2$, $\lambda=4$, $\omega_k\in[0,1]$.}
  \label{fig:2_4_lowfreq_plain}
 \end{subfigure} \hfill
 \begin{subfigure}[b]{0.48\textwidth}
  \includegraphics[width=\textwidth]{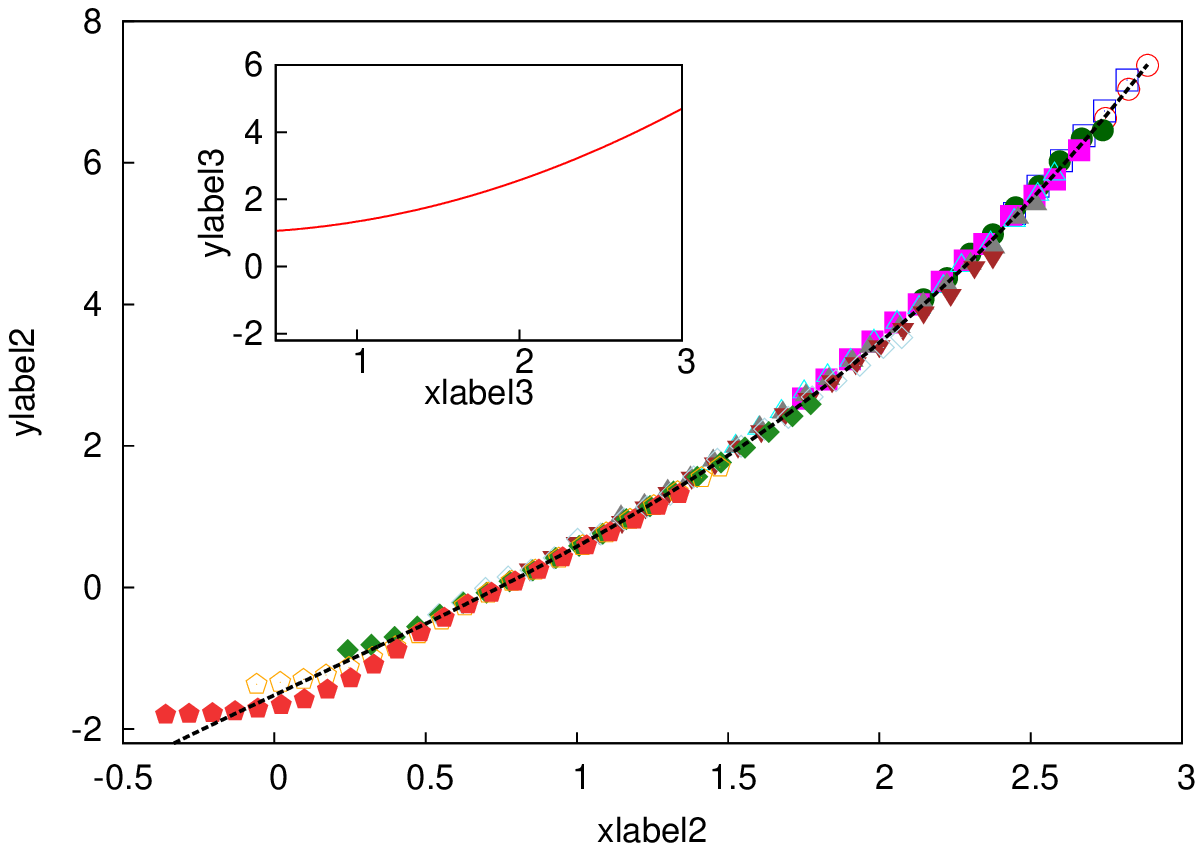}
  \caption{(b) $\Delta T(L)$ in rescaled variables.}
  \label{fig:2_4_lowfreq_scaled}
 \end{subfigure}
 % self_sim.eps: 0x0 pixel, -2147483648dpi, 0.00x0.00 cm, bb=
 \caption{Excitation times $\Delta T(L)$ for the case with harmonic on-site term and nonlinear coupling $\kappa=2$, $\lambda=4$ and on-site disorder $\omega_k\in[0,1]$.
 In panel (a) we plot the direct results $\Delta T(L)$ while in panel (b) the scaling from the FNDE with $\gamma=1$ has been applied, hence the scaled variables $\Delta T/\energy$ vs.\ $L/\energy$.
 Each color/symbol belongs to an averaged value over disorder realizations for a fixed energy $\energy$. The inlet in (b) shows the dependence of the nonlinearity index $a(w)$ on the density $w=E/L$, obtained via polynomial fitting of the data (dashed black lines).
 }
 \label{fig:2_4_lowfreq}
\end{figure}

\begin{figure}[tb]
 \centering
 \psfrag{xlabel1}[cc][cc]{\footnotesize $\log_{10}L$}
 \psfrag{ylabel1}[cc][cc]{\footnotesize $\log_{10}\Delta T$}
 \psfrag{xlabel2}[cc][cc]{\footnotesize $\log_{10}L/\energy$}
 \psfrag{ylabel2}[cc][cc]{\footnotesize $\log_{10}\Delta T/\energy$}
 \psfrag{xlabel3}[cc][cc]{\footnotesize $-\log_{10}w$}
 \psfrag{ylabel3}[cc][cc]{\footnotesize $a(w)$}
 \begin{subfigure}[b]{0.48\textwidth}
  \includegraphics[width=\textwidth]{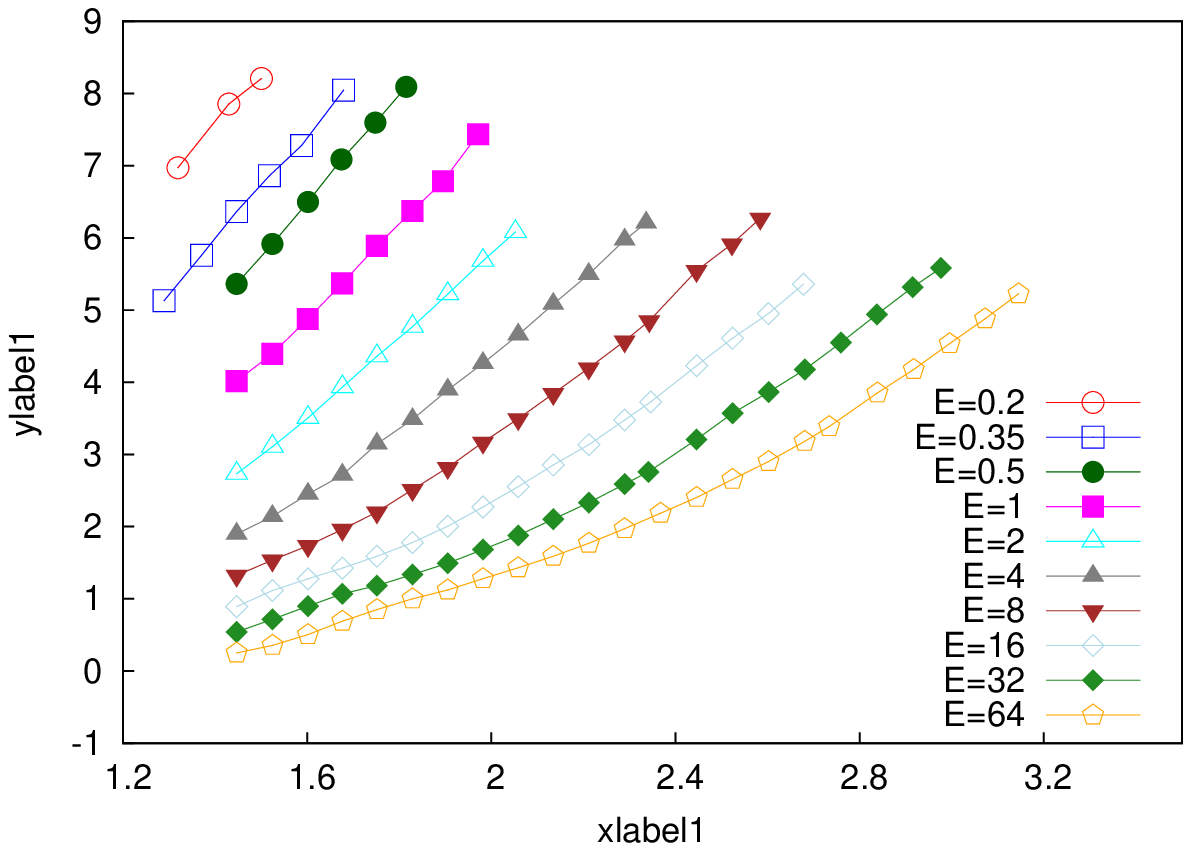}
  \caption{(a) $\Delta T(L)$ for $\kappa=2$, $\lambda=4$, $\omega_k\in[0.5,1.5]$}
  \label{fig:2_4_midfreq_plain}
 \end{subfigure} \hfill
 \begin{subfigure}[b]{0.48\textwidth}
  \includegraphics[width=\textwidth]{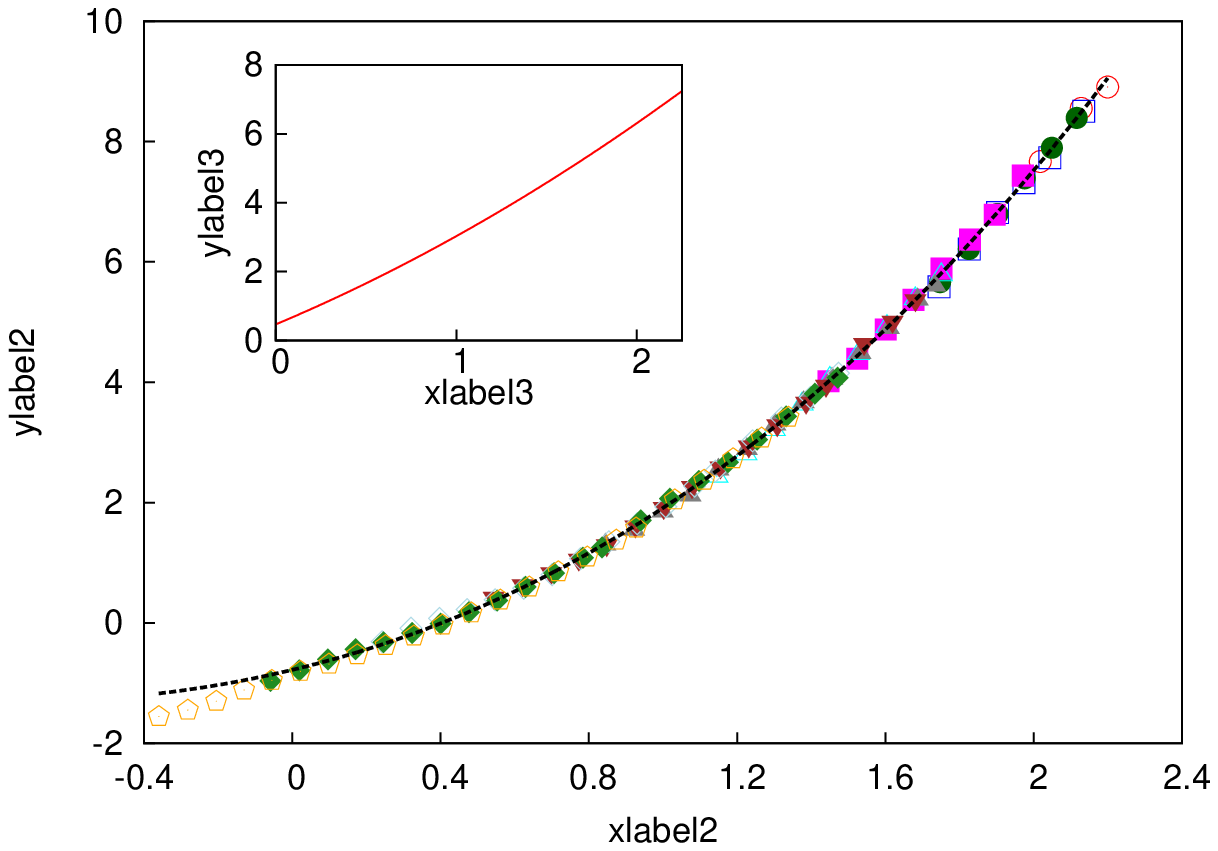}
  \caption{(b) $\Delta T(L)$ in rescaled variables.}
  \label{fig:2_4_midfreq_scaled}
 \end{subfigure}
 % self_sim.eps: 0x0 pixel, -2147483648dpi, 0.00x0.00 cm, bb=
 \caption{Excitation times $\Delta T(L)$ for the case with linear on-site term and nonlinear coupling $\kappa=2$, $\lambda=4$ and on-site disorder $\omega_k\in[0.5,1.5]$. Panel (a) shows the plain data while in (b) we applied the scaling of the FNDE with $\gamma=1$.
 The inlet in (b) shows the dependence of the nonlinearity index $a(w)$ on the density $w=E/L$, obtained via polynomial fitting of the data (dashed black lines).
 }
 \label{fig:2_4_midfreq}
\end{figure}

At first, we report the results for  $\kappa=2$ and $\lambda=4$.
Note that some of these results have already been presented in~\cite{Mulansky-Ahnert-Pikovsky-11}.
All results are again averaged over different realizations of disorder and we studied two ways of choosing the random frequencies.
%TODO:
%\todo{check M}
Figure~\ref{fig:2_4_lowfreq} shows the results for $\omega_k \in [0,1]$ and Figure~\ref{fig:2_4_midfreq} for $\omega_k \in [0.5,1.5]$.
Both cases are qualitatively very similar.
At first, we identify the energy scaling to seemingly follow the prediction of the FNDE with $\gamma=1$ as seen from the good overlaps in Figures~\ref{fig:2_4_lowfreq_scaled} and~\ref{fig:2_4_midfreq_scaled}.
The resulting curves, however, are not  straight lines but rather exhibit a clear curvature bending upwards.
This has already been reported earlier~\cite{Mulansky-Ahnert-Pikovsky-11} and is not yet fully understood.
Phenomenologically this behavior can be quantified by introducing a density dependent nonlinearity index $a$:
\[
a(w)=\frac{d \log \frac{\Delta T}{\energy}}
{d\log \frac{L}{\energy}}-1\;.
\]
From \eqref{eqn:FNDE_dT} one finds that for $\gamma=1$ the slope of the rescaled curves is simply given by $a(w)+1$.
Thus we evaluate this slope by means of a polynomial fit and plot the resulting numerical value for $a(w)$ in the insets in Figures~\ref{fig:2_4_lowfreq_scaled} and~\ref{fig:2_4_midfreq_scaled}.
Qualitatively, there is no difference between the two choices of disorder in Figures~\ref{fig:2_4_lowfreq} and~\ref{fig:2_4_midfreq}, but quantitatively the increase of the nonlinearity index $a(w)$ is faster for $\omega_k \in [0.5,1.5]$.

\begin{figure}[tb]
 \centering
 \psfrag{xlabel1}[cc][cc]{\footnotesize $\log_{10}L$}
 \psfrag{ylabel1}[cc][cc]{\footnotesize $\log_{10}\Delta T$}
 \psfrag{xlabel2}[cc][cc]{\footnotesize $\log_{10}L/\energy$}
 \psfrag{ylabel2}[cc][cc]{\footnotesize $\log_{10}\Delta T/\energy$}
 \psfrag{xlabel3}[cc][cc]{\footnotesize $-\log_{10}w$}
 \psfrag{ylabel3}[cc][cc]{\footnotesize $a(w)$}
 \begin{subfigure}[b]{0.48\textwidth}
  \includegraphics[width=\textwidth]{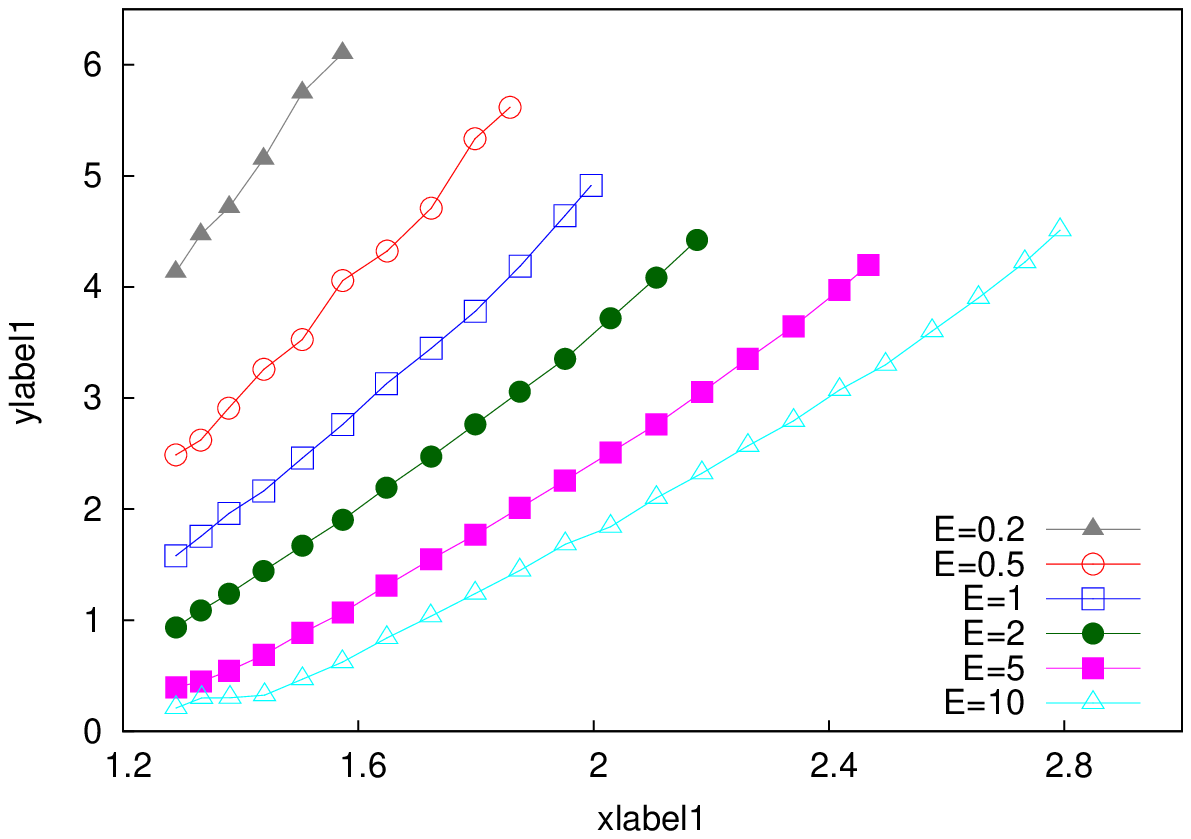}
  \caption{(a) $\Delta T(L)$ for $\kappa=2$, $\lambda=6$, $\omega_k\in[0,1]$.}
  \label{fig:2_6_lowfreq_plain}
 \end{subfigure} \hfill
 \begin{subfigure}[b]{0.48\textwidth}
  \includegraphics[width=\textwidth]{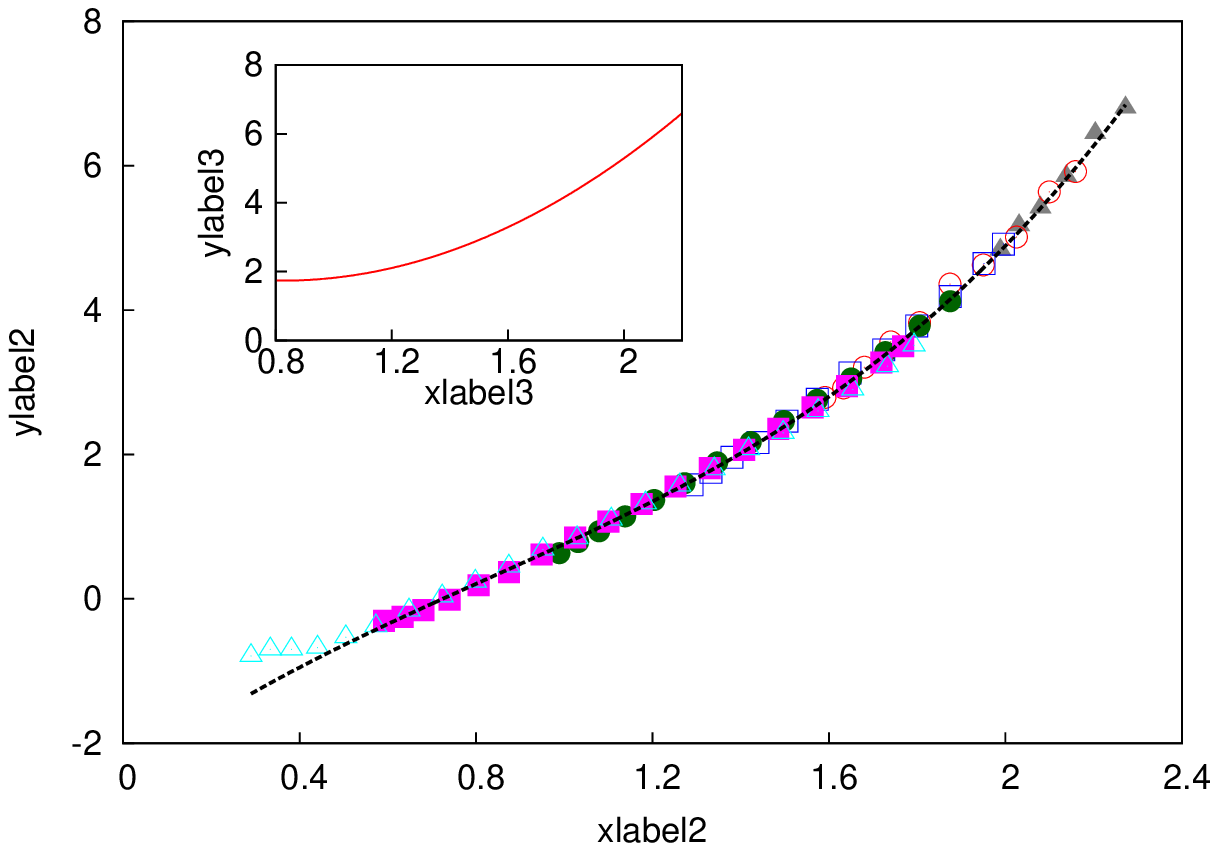}
  \caption{(b) $\Delta T(L)$ in rescaled variables.}
  \label{fig:2_6_lowfreq_scaled}
 \end{subfigure}
 % self_sim.eps: 0x0 pixel, -2147483648dpi, 0.00x0.00 cm, bb=
 \caption{Excitation times $\Delta T(L)$ for the case with harmonic on-site term and nonlinear coupling $\kappa=2$, $\lambda=6$ and on-site disorder $\omega_k\in[0,1]$.
 In panel (a) we plot the direct results $\Delta T(L)$ while in panel (b) the scaling from the FNDE with $\gamma=1$ has been applied, hence the scaled variables $\Delta T/\energy$ vs.\ $L/\energy$.
 Each color/symbol belongs to an averaged value over disorder realizations for a fixed energy $\energy$. The inlet in (b) shows the dependence of the nonlinearity index $a(w)$ on the density $w=\energy/L$, obtained via polynomial fitting of the data (dashed black lines).
 }
 \label{fig:2_6_lowfreq}
\end{figure}

\begin{figure}[tb]
 \centering
 \psfrag{xlabel1}[cc][cc]{\footnotesize $\log_{10}L$}
 \psfrag{ylabel1}[cc][cc]{\footnotesize $\log_{10}\Delta T$}
 \psfrag{xlabel2}[cc][cc]{\footnotesize $\log_{10}L/\energy$}
 \psfrag{ylabel2}[cc][cc]{\footnotesize $\log_{10}\Delta T/\energy$}
 \psfrag{xlabel3}[cc][cc]{\footnotesize $-\log_{10}w$}
 \psfrag{ylabel3}[cc][cc]{\footnotesize $a(w)$}
 \begin{subfigure}[b]{0.48\textwidth}
  \includegraphics[width=\textwidth]{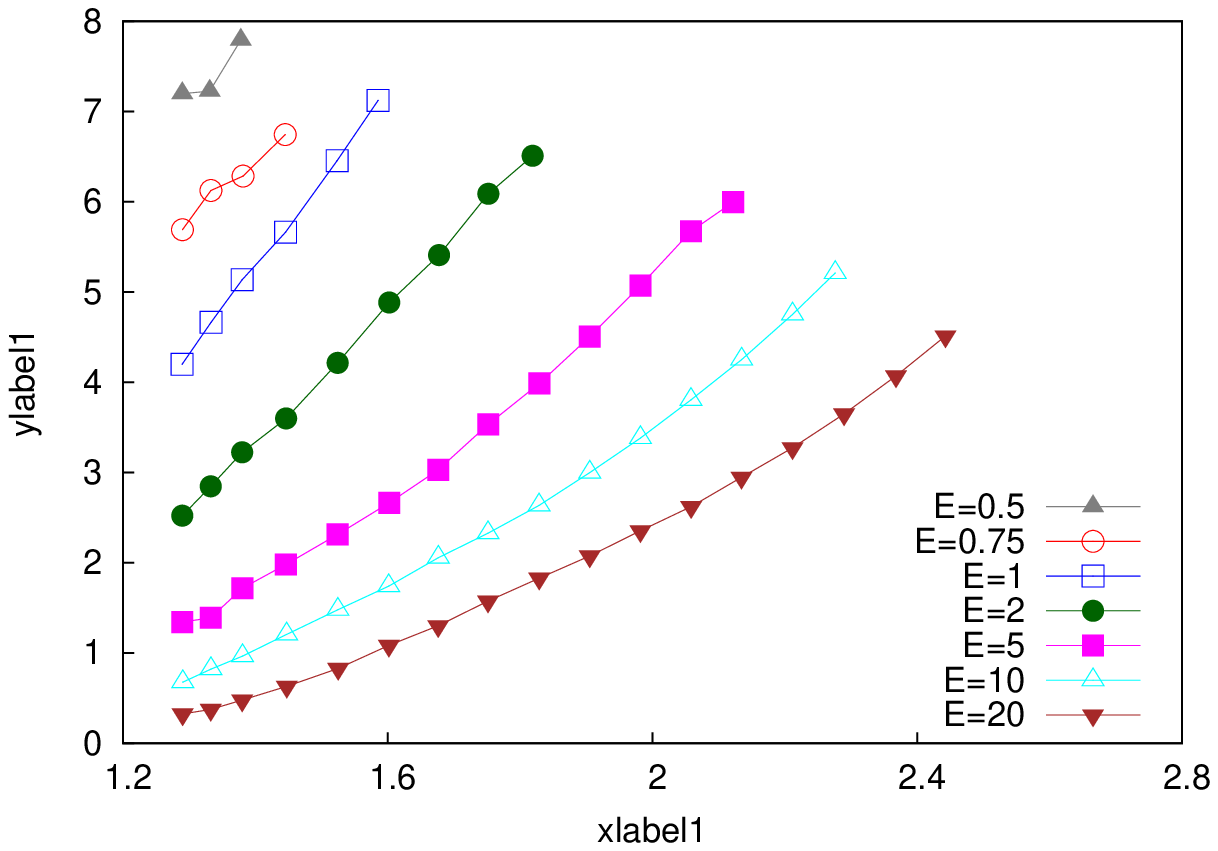}
  \caption{(a) $\Delta T(L)$ for $\kappa=2$, $\lambda=6$, $\omega_k\in[0.5,1.5]$}
  \label{fig:2_6_midfreq_plain}
 \end{subfigure} \hfill
 \begin{subfigure}[b]{0.48\textwidth}
  \includegraphics[width=\textwidth]{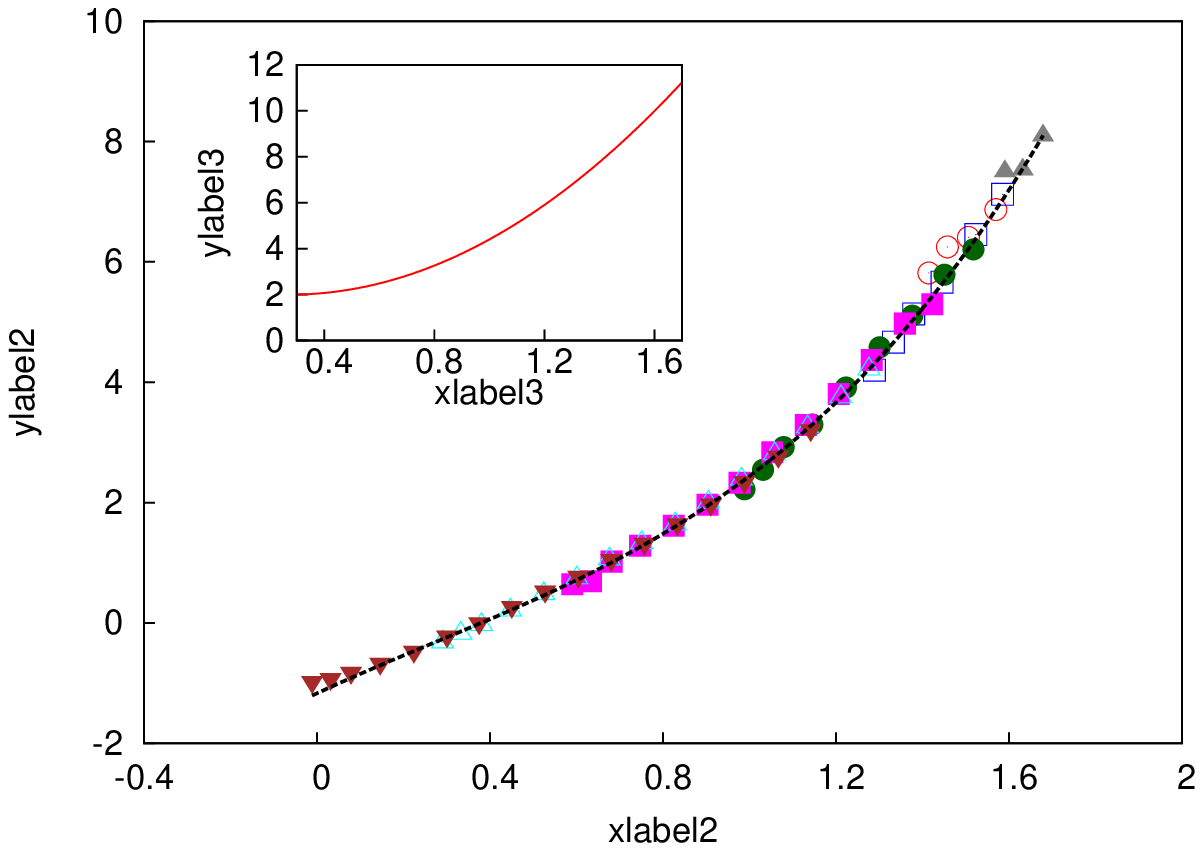}
  \caption{(b) $\Delta T(L)$ in rescaled variables.}
  \label{fig:2_6_midfreq_scaled}
 \end{subfigure}
 % self_sim.eps: 0x0 pixel, -2147483648dpi, 0.00x0.00 cm, bb=
 \caption{Excitation times $\Delta T(L)$ for the case with linear on-site term and nonlinear coupling $\kappa=2$, $\lambda=6$ and on-site disorder $\omega_k\in[0,1]$. Panel (a) shows the plain data while in (b) we applied the scaling of the FNDE with $\gamma=1$.
 The inlet in (b) shows the dependence of the nonlinearity index $a(w)$ on the density $w=\energy/L$, obtained via polynomial fitting of the data (dashed black lines).
 }
 \label{fig:2_6_midfreq}
\end{figure}

In Figures~\ref{fig:2_6_lowfreq} and~\ref{fig:2_6_midfreq} we show the results of a similar study with the coupling nonlinear exponent $\lambda=6$.
The results are again qualitatively the same as above in that we find scaling with $\gamma=1$ and a density dependent nonlinearity index $a(w)$ shown in the insets of Figures~\ref{fig:2_6_lowfreq_scaled} and~\ref{fig:2_6_midfreq_scaled}.
Hence, this seems to be a universal picture for spreading in lattices of harmonic oscillators with random frequencies and nonlinear nearest neighbor coupling.
It should be noted that the density dependent spreading can not be described by introducing a density dependent parameter of the fractional derivative $\gamma(w)$, because this would mean a density dependent energy
scaling which is not observed here.
We also note that by introducing a density dependent nonlinearity index $a(w)$ into the FNDE (or NDE as we have $\gamma=1$ here) destroys the self-similar solution and even the scaling prediction.
However, the density dependence is found to be very weak $a(w)\sim \log_{10} w$ and thus the rate of change of $a$ is much slower then the spreading time scale.
Thus, it is reasonable to treat the energy spreading in a first approximation using $a=\text{const}$ and then analyze the slow deviations afterwards.
The question of the asymptotic behavior remains, however, open: from the data presented here we cannot judge whether the spreading effectively stops, or continues with an increasing index $a$, or some transition to another law of spreading (e.g., a logarithmic one) occurs.

\section{Conclusions}

Motivated by previous observations of subdiffusive behavior in nonlinear disordered systems and anomalous diffusion in chaotic Hamiltonian systems, we introduced the fractional nonlinear diffusion equation as a phenomenologic model to describe the spreading process in disordered one-dimensional Hamiltonian lattices of nonlinearly coupled oscillators.
We have found that with the FNDE it is possible to explain in a consistent way  the subdiffusive spreading behavior and the energy scaling of spreading states.
Analysis of self-similar solutions of the FNDE not only predicts a subdiffusive spreading, but also induces a scaling of time and energy of the spreading process according to relations (\ref{eqn:FNDE_spreading}, \ref{eqn:FNDE_dT}), which depend on parameters $\gamma$ and $a$, responsible for the index of the fractional time derivative and of the nonlinearity, respectively.
We tested these scaling laws on a class of nonlinearly coupled oscillators with different values of the nonlinear indices $\kappa$ (local nonlinearity)  and $\lambda$ (coupling nonlinearity).
Our main result is that there are three qualitatively different ``universality classes'' in regard of relations between $\gamma,a$ and $\kappa,\lambda$.  Specifically, we have found the following three cases of nonlinearities that demonstrate different scaling of spreading:

(i) For homogeneous nonlinear potentials, where $\kappa=\lambda$, we were able to deduce an exact spreading prediction from the scaling property of the Hamiltonian equations and the FNDE when assuming a fully chaotic phase space.
We argued  that here the nonlinear diffusion equation with
$\gamma=1$, i.e. with normal time derivative and the nonlinearity index $a=\frac{\kappa-2}{2\kappa}$ should be applied.
This analytic prediction has been confirmed numerically as the asymptotic spreading behavior.
As an important result we again note that subdiffusive spreading was also found in the regular case without disorder.
This further supports the claim that disorder is not required for the spreading and it indeed seems reasonable to call this process ''chaotic diffusion``~\cite{Mulansky-Pikovsky-12,Mulansky_phd}.
%In all considered cases the one-parameter scaling laws were nicely reproduced in our numerical study. 

(ii) In the fully nonlinear case with local nonlinearity index of the oscillators $\kappa=4$ and the nonlinearity indexes $\lambda=6,8$ in the coupling, we have found that the numerical spreading results follow the energy scaling as predicted from the FNDE with the fractional time derivative of order  $\gamma=1.08$ (for $\lambda=6$) and $\gamma=1.18$ (for $\lambda=8$).
This is compatible with previous findings on anomalous diffusion in low-dimensional Hamiltonian systems were the mixed phase space also leads to a fractional diffusion equation with $\gamma>1$~\cite{Zaslavsky-05}.
Furthermore, for this case we were able to construct a microscopic model of the dynamics at the excitation edge that predicts the correct spreading behavior verified in direct numerical simulations.

(iii) In the case of nonlinearly coupled harmonic disordered oscillators, we have verified that the energy scaling  follows nicely the prediction of the normal nonlinear diffusion equation (fractional order $\gamma=1$).
However, the spreading does not follow a pure power law as predicted by the NDE. Instead, we have identified a remarkable dependence of the effective index of nonlinearity of the FNDE on the energy density $a(w)$. In all cases considered we have observed that $a$ increases as $w$ becomes smaller, although the particular profiles of $a(w)$ depend on the nonlinearity in coupling and on the disorder.
As the effective nonlinearity increases in the course of spreading, this means a slowing down of the spreading process compared with the perfect power law,  as in this case $L \sim t^{\frac{1}{a(w)+2}}$.
Unfortunately, we are not able to present a microscopic model of the edge dynamics at this point, mainly due to the highly complicated resonance structure that emerges when considering nonlinearly coupled harmonic oscillators with random frequencies.
Consequently, it is also not possible to judge from the data what is the asymptotic behavior of the spreading for times beyond those available in our numerics.

Our findings rely to a large extent  on the novel quantity characterizing the spreading, the averaged excitation time introduced in \cite{Mulansky-Ahnert-Pikovsky-11}. 
This quantity is defined for a particular size of the wave packet, and thus for a particular value of the density.
It thus allows us to reveal the density dependence of the spreading characteristics, what is hardly possible with old approaches where, e.g., averaged participation numbers have been followed. 
Unfortunately, the calculation of averaged excitations is relying on the sharp edges of the field, so its application to linearly coupled lattices where eigenmodes are exponentially (but not sharp) localized, remains a challenge for future studies. We stress once more that in our study we consider the fractional nonlinear diffusion equation as a phenomenological model guiding the scaling relations of the problem. Its derivation from the microscopic model appears, at the present stage, as a complex, not yet resolved problem. In this respect we refer to paper~\cite{Schwiete-Finkelstein-10}, where an attempt to derive a nonlinear diffusion equation for the two-dimensional disordered nonlinear Schr\"odinger equation is presented; the resulting conclusion on the linear in time growth of the variance of the wave packet (like in normal diffusion) does not, however, correspond to numerical findings of subdiffusion in this model~\cite{Garcia-Mata-Shepelyansky-09a}. Further attempts are necessary to resolve this problem.

\section*{Acknowledgement}
The numerical results have been partially obtained at the CINECA sp6 supercomputer under the Project HPC-EUROPA2 (Project number 228398), with the support of the European Community - under the FP7 ``Research Infrastructure'' Programme.
M.~M.\ thanks the CNR Institute for Complex Systems in Florence and the IHP Paris for hospitality and financial support, and DFG for support under project \mbox{PI 220/12-1}. 
Fruitful discussions with D. Shepelyansky, S. Fishman, and S. Flach are cordially acknowledged.

%\bibliographystyle{unsrt}
% \bibliography{nld-old,nld-current,%
% pap-ab,pap-ce,pap-fg,pap-hj,pap-kl,%
% pap-mn,pap-oq,pap-rs,pap-tz,%
% pik,books,n-stand}

\end{document}